%% file: lcsms.tex
\documentclass[fleqn,usenatbib]{mnras}

\usepackage{newtxtext,newtxmath}

\usepackage[T1]{fontenc}
\usepackage{soul}

\DeclareRobustCommand{\VAN}[3]{#2}
\let\VANthebibliography\thebibliography
\def\thebibliography{\DeclareRobustCommand{\VAN}[3]{##3}\VANthebibliography}

\usepackage{graphicx}	
\usepackage{amsmath}	

\usepackage{orcidlink}
\usepackage{xcolor}

\newcommand{\lcs}{$LCS$}
\newcommand{\kms}{$\rm km~s^{-1}$}
\newcommand{\sers}{S\'{e}rsic}

\newcommand{\mlstar}{$M_\star/L_r$}
\newcommand{\dsfr}{$\rm \Delta \log_{10} SFR$}
\newcommand{\bt}{$B/T$}

\title[LCS SFR-Mass Relation]{The Local Cluster Survey II: Disk-Dominated Cluster Galaxies with Suppressed Star Formation}

\author[Rose A. Finn et al.]{Rose A. Finn$^{1}$\orcidlink{https://orcid.org/0000-0001-8518-4862}\thanks{E-mail: rfinn@siena.edu (RAF)},
Benedetta Vulcani$^{2}$\orcidlink{0000-0003-0980-1499},
Gregory Rudnick$^{3}$,
Michael L. Balogh$^{4,5}$,
Vandana Desai$^{6}$,
\newauthor
Pascale Jablonka$^{7,8}$
and Dennis Zaritsky$^{9}$
\\
$^{1}$Siena College, 515 Loudon Rd.,
Loudonville, NY 12211\\
$^{2}$INAF-Osservatorio astronomico di Padova, Vicolo Osservatorio 5, I-35122 Padova, Italy\\
$^{3}$The University of Kansas, Department of Physics and Astronomy, Malott Room 1082, 1251 Wescoe Hall Drive, Lawrence, KS, 66045, USA\\
$^{4}$Department of Physics and Astronomy, University of Waterloo, Waterloo, Ontario N2L 3G1, Canada\\
$^{5}$Waterloo Centre for Astrophysics, University of Waterloo, Waterloo, Ontario, N2L3G1, Canada\\
$^{6}$IRSA, California Institute of Technology, MS 220-6, Pasadena, CA 91125, USA\\
$^{7}$Laboratoire d'astrophysique, \'Ecole Polytechnique F\'ed\'erale de Lausanne (EPFL), 1290 Sauverny, Switzerland\\
$^{8}$GEPI, Observatoire de Paris, Universit\'e PSL, CNRS, Place Jules Janssen, F-92190 Meudon, France \\
$^{9}$The University of Arizona, 933 N. Cherry Ave, Tucson, AZ, USA
}

\date{Accepted XXX. Received YYY; in original form ZZZ}

\pubyear{2022}

\begin{document}
\label{firstpage}
\pagerange{\pageref{firstpage}--\pageref{lastpage}}
\maketitle

\begin{abstract}
We investigate the role of dense environments in suppressing star formation by studying $\rm \log_{10}(M_\star/M_\odot) > 9.7$ star-forming galaxies in nine clusters from the {\it Local Cluster Survey} ($0.0137 < z < 0.0433$) and a large comparison field sample drawn from the Sloan Digital Sky Survey.  
 We compare the star-formation rate (SFR) versus stellar mass relation as a function of environment and morphology.   
After carefully controlling for mass, we find that in all environments, the degree of SFR suppression increases with increasing  bulge-to-total (\bt) ratio. 
  In addition, the SFRs of cluster and infall galaxies at a fixed mass are more suppressed than their field counterparts at all values of \bt.  These results suggest a quenching mechanism that is linked to bulge growth that operates in all environments and an additional mechanism that further reduces the SFRs of galaxies in dense environments.
  We limit the sample to $B/T \le 0.3$ galaxies to control for the trends with morphology and find that the excess population of cluster galaxies with suppressed SFRs persists.  
We model the timescale associated with the decline of SFRs in dense environments and find that the observed SFRs of the cluster core galaxies are consistent with a range of models including: a mechanism that acts slowly and continuously over a long ($2-5$~Gyr) timescale, and a more rapid ($<1$~Gyr) quenching event that occurs after a delay period of $1-6$~Gyr. Quenching may therefore start immediately after galaxies enter clusters.

\end{abstract}




\section{Introduction} \label{sec:intro}
Foremost among the results of galaxy surveys over the last decade 
has been the realization that the galaxy population at $z \lesssim 2$
is bimodal in nature \citep[][]{Strateva2001,Bell2004,Cooper2006}. 
That is, galaxies 
are effectively described as one of two distinct types: red, early-type galaxies lacking much star formation; and blue, late-type galaxies with active star formation.  There is clear evidence that galaxies transform between these types, as the cosmic stellar mass density that is located within quiescent galaxies has doubled since $z\sim 1$ 
\citep[e.g.][]{Bell2004,Bundy2006,Faber2007}.  
Despite this secure evidence for galaxy transformation, little detail is known about how the transformation of galaxies from star forming to quiescent actually occurs.

The goal of this paper is to investigate the role that dense environments play in quenching infalling galaxies using a sample of nine low-redshift clusters from the {\it Local Cluster Survey}  \citep[\lcs;][]{Finn2018}.  The clusters span a range in cluster mass from Coma, the most massive cluster in the local universe, to less massive clusters and groups.  We thus sample the full range of dense environments, and we investigate galaxies in both the core and infall regions of each cluster.  In \citet{Finn2018}, we showed that the size of the dust disk, as traced by 24\micron \ emission, relative to the stellar disk, as traced by the $r$-band emission, is lower in the cluster cores than in the infall regions.  In this paper, we focus on galaxy star-formation rates (SFRs).  

When isolating the impact of environment on star formation, we must carefully consider that the amount of star formation in galaxies is strongly correlated with stellar mass, morphology, and environment \citep[e.g.][]{Kennicutt1998, Peng2010, Poggianti2006,Poggianti2008,Vulcani2010}.   The correlation between stellar mass and quenching is well documented, with the fraction of quiescent galaxies increasing with stellar mass \citep[e.g.][]{Blanton2009,Peng2010}.  Several mechanisms could drive this trend, including feedback from an active galactic nucleus (AGN) and massive star formation, as both can inject energy into galactic gas.  The gas is heated, which can prevent subsequent collapse and thus star formation, or the gas can be completely ejected from the galaxy by the strong winds \citep[e.g.][]{DeLucia2004, Oppenheimer2008, Abramson2014,Muratov2015}.  
 While the fraction of star-forming galaxies decreases with increasing stellar mass, the amount of star formation within star-forming galaxies increases with increasing stellar mass. 
This dependence is frequently quantified using the SFR-Stellar Mass relation  \citep[e.g.][]{Elbaz2007, Noeske2007, Salim2007}.
The strong correlation between star formation and stellar mass persists out to high redshift \citep{Whitaker2012, Speagle2014,Schreiber2016}, although the normalization changes due to the cosmic evolution of star formation \citep[e.g.][]{Lilly1996, Madau1996}.

The second parameter that is strongly correlated with star formation is galaxy morphology.  Most obviously, there is little to no star formation in ellipticals and S0s, whereas spiral galaxies host significant amounts of star formation \citep[e.g.][]{Kennicutt1998}.  However, even within spiral galaxies, the amount of star-formation is correlated with morphology, with late-type spirals having higher specific star formation rates (sSFR; SFR divided by stellar mass) than early-type spirals \citep[e.g.][]{Kennicutt1998}.  The correlation between SFR and morphology persists when more quantitative metrics are used to characterize morphology such at $B/T$ or concentration \citep[e.g.][]{Zhang2021}. 
One explanation for this correlation is that the processes that build bulges, such as galaxy-galaxy interactions, can funnel gas to the center, thereby consuming the fuel through star formation \citep[e.g.][]{Mihos1996}, although studies of field post-starburst galaxies show that the merging process might lead to lower SFRs by changing the state of the gas rather than by depleting it \citep[e.g.][]{French2015}.   Other processes that are correlated with the size of the bulge can lead to quenching without depleting the gas.  For example, in morphological quenching the bulge can stabilize the disk against gravitational collapse and reduce star formation \citep[][]{Martig2009}.  


{Dense environments are expected to affect the rate at which galaxies form stars and eventually induce their quenching by altering their gas reservoir through hydrodynamic stripping of the extended 
or disk gas.  
Ram pressure stripping of the disk gas due to the interaction between the galaxy interstellar medium (ISM) and the intergalactic medium \citep[IGM,][]{Gunn1972} is one of the most efficient means of 
removing the ISM, leaving a recognizable pattern of star formation with truncated H$\alpha$ disks smaller than the undisturbed stellar disk \citep[e.g.][]{Koopmann2004,Yagi2015, Bell2004}
and truncated gas disks \citep[e.g.][]{Dale2001,Koopmann2004, Cortese2010, Boselli2016, Finn2018}. 
On local scales, ram-pressure stripping is expected to operate on short time scales, of the order of a Gyr (e.g. \citeauthor{Quilis2000} \citeyear{Quilis2000}) even though globally timescales associated with ram-pressure stripping can be much longer \citep{Tonnesen2019}. 
Strangulation is the removal of the hot gas halo surrounding the galaxy either via ram pressure or via tidal stripping by the halo potential \citep{Larson1980,Balogh2000}.
This deprives the galaxy of its gas reservoir but leaves the existing ISM in the disk to be consumed by star formation on timescales of 2-3 Gyr \citep[e.g.][]{Bigiel2008}, without leaving any clearly asymmetric features as in the case of ram pressure stripping. Strong tidal interactions and mergers, and tidal effects of the cluster as a whole, can also deplete the gas in an inhomogeneous way, similarly affecting both the gas and stellar component and leaving a visible signature that persists for few Gyr. 
}


{The time-scale of the transition from being actively star-forming galaxies to becoming passive in these scenarios must be completely different. 
Different models are being developed to reconcile conflicting observations that suggest quenching occurs on either fast
\citep[$<1$ Gyr; e.g.][]{Balogh2004,McGee2009,Muzzin2012}
or slow  \citep[2-4 Gyr; e.g.][]{Wolf2005,Vulcani2010,DeLucia2012,Haines2013,Wheeler2014,Reeves2022} timescales. 
One of the most popular models in the one proposed by \citet{Wetzel2013}, which posits that galaxies enter a more massive halo and resides there for a ``delay time" before undergoing a rapid quenching event. According to their results, satellite SFRs evolve unaffected for 2–4~Gyr after infall, after which star formation quenches rapidly, with an e-folding time of $<$0.8 Gyr \citep[see also ][]{Haines2013,Phillipps2019, Wright2019,Rhee2020}. Quenching time-scales are shorter for more massive galaxies but do not depend on halo mass of the hosting system: the observed increase in the satellite quiescent fraction with halo mass arises simply because satellites quench in a lower mass group prior to infall (group preprocessing), which is responsible for up to half of quenched satellites in massive clusters.
This "delay+rapid" model has been consistently supported by many studies due to its suitability for describing the quenching of galaxies 
\citep{Wetzel2013,McGee2014,Mok2013,Tal2014,Balogh2016,Fossati2017,Foltz2018}.
A rapid quenching event is needed to explain post-starforming/post-starburst galaxies, which are galaxies with spectral properties that suggest quenching occurred very fast and in a recent past
\citep[e.g.][]{Poggianti1999,Poggianti2004,Poggianti2009,Paccagnella2016,Vulcani2020}.\footnote{Note though that in clusters some mechanisms can also provoke a temporary enhancement of the star formation prior to quenching \citep[e.g.][]{Moss1993, Poggianti2016, Vulcani2018, Roberts2020}.}
}

 {The arguments in favor of slow versus rapid quenching hinge on the presence (or lack thereof) of an excess population of galaxies with suppressed star-formation in dense environments. Much of the apparent contradiction in observational results can be explained by differences in how the samples are selected and how environment is parameterized \citep[e.g.][]{Muldrew2012}. For example, studies that impose a high threshold on observed SFRs \citep[e.g.][]{Finn2008,Verdugo2008,Bamford2008,Peng2010} or emission line equivalent widths \citep[e.g.][]{Balogh2004} will not be sensitive to galaxies with suppressed SFRs.   In contrast, studies with lower SFR limits detect a population of cluster galaxies with lower SFRs than their field counterparts of similar stellar mass \citep{Vulcani2010, Finn2010, Paccagnella2016, RodriguezdelPino2017, Guglielmo2019}. In addition, some studies that investigate the role of environment do not include clusters \citep[e.g.][]{Calvi2013} or quantify environment in terms of local densities, binning in a way that washes out the dense environments corresponding to galaxy clusters \citep[e.g.][]{Peng2010}.  Thus they are not adequately probing the environments where ram-pressure stripping is most likely to occur. }

Identifying the role that environment plays in quenching galaxies and the timescale associated with quenching is complicated for several reasons.  First, 
multiple mechanisms are working to remove gas in galaxies in dense environments, and typically each experiment is sensitive to a particular range of physics.  
Second, the direct link between physical mechanisms and timescales may not be straightforward, as even processes that are thought to act quickly, like ram-pressure stripping, might act on significantly longer timescales that depend on the details of a galaxy's orbit through the cluster  \citep{Tonnesen2019}.  
Finally, perhaps the most complicating factor is that the parameters that correlate with star formation (stellar mass, morphology, and environment), are themselves correlated.  Thus, to isolate the role of environment, one must carefully control for stellar mass and morphology \citep[e.g.][]{Weinmann2009,Salmi2012,Guglielmo2015,Morselli2017,Lofthouse2017,Spindler2018,Liu2019, Sampaio2022}.  

{The goal of this study is to isolate the effect of the cluster environment on galaxy star-formation rates.}  We will compare the star-forming cluster galaxies to a large sample of field galaxies (samples described in Section \ref{sec:sample}), and we will quantify the SFRs while carefully controlling for stellar mass and morphology.  Importantly, we probe to low stellar mass ($\log_{10}(M_\star/M_\odot) > 9.7)$, where environmental quenching at low redshift is expected to be dominant, and to low SFR, which enables us to detect galaxies with suppressed star formation.
We use the SFR-mass relation of galaxies in the cluster core and infall regions, in comparison to that in the field, to identify an excess population of galaxies with suppressed SFRs in dense environments (Section \ref{sec:results}). In the second part of the paper, we take into account the impact of morphology on the SFR-mass relation by excluding bulge-dominated galaxies -- which have been shown to drive the sSFR-mass correlation \citep{Abramson2014}  --  and find that the population of cluster galaxies with suppressed SFRs persists even among the disk-dominated sample (Section \ref{sec:disk_restrict}).  
We also develop a tool to quantify the timescale of the observed SFR decline (Section \ref{sec:model}).
%
We discuss the implications of our results and modeling in Section \ref{sec:discussion} and present our conclusions in Section \ref{sec:conclusions}.  We assume WMAP-9 cosmology and \citep{Chabrier2003} IMF  throughout.

\section{Galaxy Samples and Properties}\label{sec:sample}

\subsection{{\it Local Cluster Survey} }
The Local Cluster Survey (\lcs) is a {\it Spitzer Space Telescope} wide-area survey of nine $0.0137 < z < 0.0433$ clusters.  The cluster and galaxy selection are described in detail in \citet{Finn2018}, and here we just report the most important aspects of the selection. 
All of the \lcs\ clusters lie within the SDSS \citep{York2000} survey, and they were selected to span a wide range of velocity dispersions ($300 < \sigma  < 1100$~\kms), X-ray luminosity (L$_X =$0.1–2.4keV), and X-ray temperature so that they probe the full range of intra-cluster medium properties.  
All of the clusters have wide-area $Spitzer$ MIPS 24$\mu$m mapping taken either from the $Spitzer$ Science Archive or obtained specifically for this project.  In this paper, we do not utilize the $Spitzer$ data, and so we do not need to restrict the sample to galaxies that fall within the $Spitzer$ footprint as we did in \citet{Finn2018}.  
As in \citet{Finn2018}, we use the the NASA-Sloan Atlas as the parent catalog \citep{Blanton2011}.

\subsubsection{Definition of Environments} \label{sec:envs}
We define the cluster core and infall regions using a  projected phase-space diagram that  
relates the velocity of the galaxies relative to the cluster velocity dispersion ($\sigma$) and the galaxy clustercentric distance, normalized by the cluster $R_{200}$.  The values of $\sigma$ and $R_{200}$ are calculated from the cluster biweight scale, and we take the values from \citet{Finn2018}.
The sample is divided into galaxies in the cluster core and the infalling regions using a phase-space cut from \citet{Oman2013}. Specifically, galaxies in the region defined by $|\Delta v/\sigma|< -4/3\times \Delta R/R_{200} + 2$ 
 are likely to be true cluster members (from now on {\it core galaxies}). The region outside the \cite{Oman2013} cut
contains  galaxies near the cluster as well as a large fraction ($>$50\%) of interlopers that are not physically associated with the cluster \citep{Oman2013}. To exclude such interlopers, we limit to $\Delta v/\sigma < 3\sigma$ and $\Delta R < 3R_{200}$. Galaxies within this cut are likely to lie near the cluster, and we refer to this population as  {\it infall} galaxies. Galaxies with  $\Delta v/\sigma > 3\sigma$ are likely external to the cluster and will be disregarded in this analysis.
We  point out that  there may be a significant contribution from backsplash  galaxies in the {infall} region, i.e. those galaxies that have already crossed $R_{200}$ at least once \citep[e.g.][]{Balogh2000,Oman2013}.  However, this paper focuses on star-forming galaxies (\S\ref{sec:galaxy_properties}), and it is not clear what fraction of star-forming galaxies in the {infall} region are indeed backsplash galaxies, as past estimates of the fraction of backsplash galaxies in this region were computed for galaxies irrespective of their star-formation histories \citep{Balogh2000}.  Regardless, we must therefore keep in mind that our {infall} region is not purely made up of freshly infalling galaxies.  Furthermore, both the core and infall samples will be contaminated by interloper galaxies that are not physically associated with the cluster.  Any such interlopers will serve to wash out intrinsic differences between the field and cluster/infall environments.

\subsection{Field Sample}
We assemble an appropriate field comparison sample making use 
of the environmental catalog from \citet{Tempel2014}, which is created from the DR10 Sloan Digital Sky Survey (SDSS) spectroscopic sample ($r \leq 17.77$). The authors used a modified friends-of-friends (FoF) method with a variable linking length in the transverse and radial directions to find as many groups as possible, while keeping the general group properties uniform with respect to distance  \citep[see][for further details]{Nurmi2013, Old2014}. 
For each galaxy in the catalog, \citet{Tempel2014} provide a halo mass estimate of the hosting structure.  We create our field sample from galaxies that (1) belong to halos with $\log_{10} (M_{halo}/M_\odot) < 13$, and (2) lie in the same redshift range as the \lcs \ galaxies. To be clear, our field sample is defined differently from many observational studies in which the field is used to approximate the galaxy properties in a typical volume of the universe, so long as that volume does not contain a cluster.  Thus, more typical field samples could include massive groups, whereas ours will not.  We test the affect of our selected halo mass limit and find that a slightly different choice of the field sample (e.g. $\log_{10} (M_{halo}/M_\odot) < 12.5$ or $\log_{10} (M_{halo}/M_\odot) < 13.5$) does not affect our results.

\subsection{Galaxy properties}
\label{sec:galaxy_properties}
\begin{figure}
    \centering
    \includegraphics[width=0.5\textwidth]{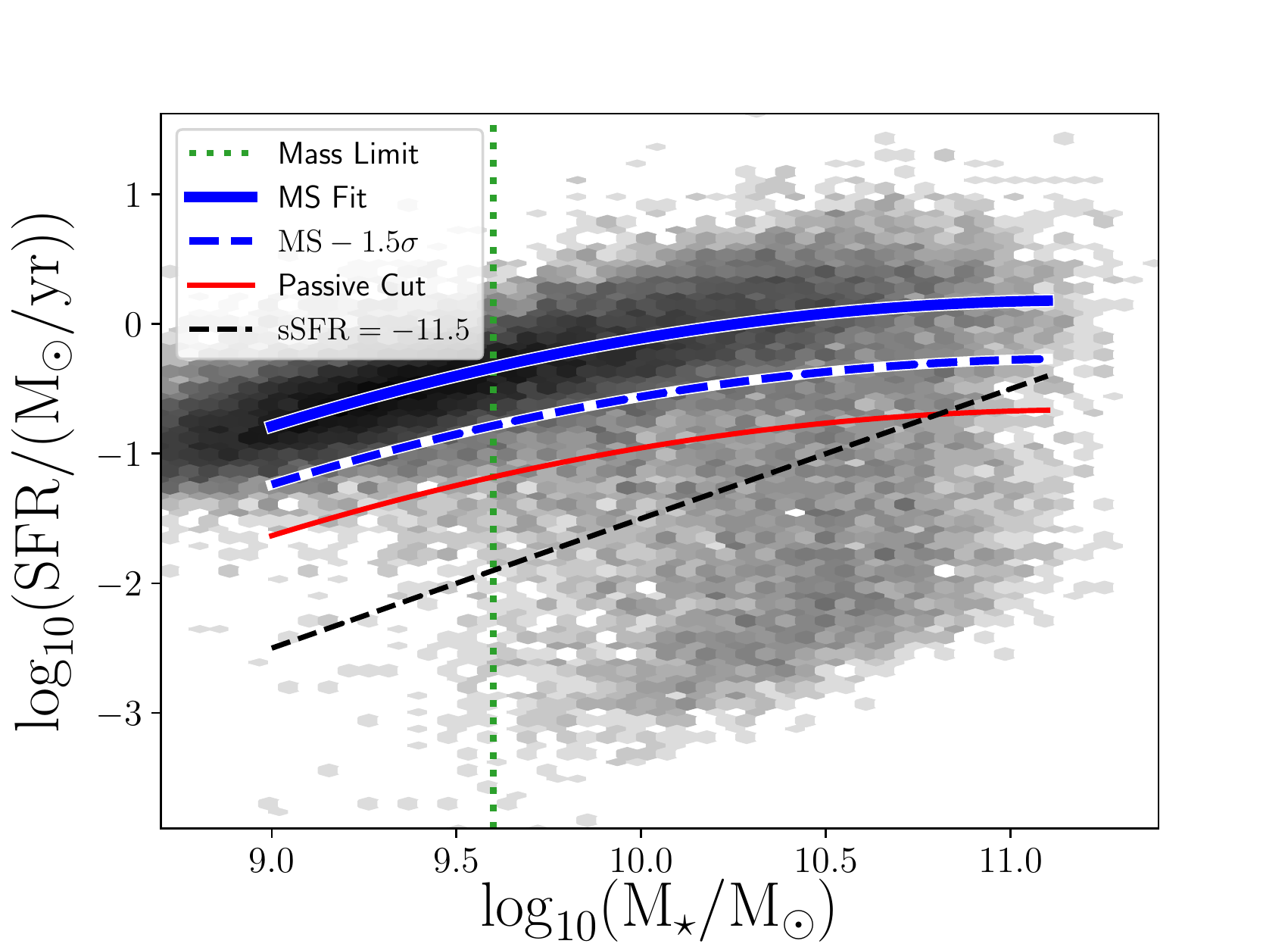}
    \caption{SFR versus stellar mass for the full field sample.  The blue line is the best fit to the main sequence (see Appendix \ref{append:sfms_fit} for details on the fit), and the dashed blue line corresponds to the main sequence fit minus 1.5 times the dispersion in the fit.  The passive galaxies begin to dominate the population approximately 0.8 dex below the main sequence, and we show this division with the red line (Appendix \ref{append:passive}).  We define our star-forming sample as galaxies that lie above the red line.  The black dashed line shows a specific SFR of $-11.5$, which is the minimum reliable value of the GSWLC-2 SFRs \citep{Salim2018}.  }
    \label{fig:ms_passive_cut}
\end{figure}

In order to compare the cluster and field samples, we need to use homogeneous SFR and stellar mass estimates. 
We match both the \lcs \ and field samples with the  GALEX-Sloan-WISE Legacy Catalog \citep[GSWLC2; ][]{Salim2016, Salim2018} using a search radius of 10\arcsec. The GSWLC2 catalog 
provides  physical properties (stellar masses, dust attenuations, and SFRs) for $\sim$700,000 galaxies with SDSS redshifts below $z < 0.3$. 
We use the GSWLC2-X catalog, which uses the deepest imaging available for each galaxy from among the deep, medium, and all-sky GALEX surveys \citep{Salim2018}.  
The uncertainty of GSWLC stellar masses is $0.03 - 0.13$ dex for the range in specific SFRs covered by our sample \citep{Salim2016}.
The SFRs are derived by fitting templates to the UV through optical SED plus infrared luminosity.  The total infrared luminosity is used to constrain the SED, without fitting the full IR SED.  \citet{Salim2018} determine the minimum reliable sSFR to be $\log_{10}(sSFR/yr^{-1}) = -11.5$.  This is the limit associated with GALEX all-sky survey, the shallowest of the three GALEX surveys. 
We use a more conservative approach for separating star-forming and passive galaxies, and we show the division using the red curve in Figure \ref{fig:ms_passive_cut}.  We describe the procedure used to determine the main sequence fit (blue line in Fig. \ref{fig:ms_passive_cut}) in Appendix \ref{append:sfms_fit} and the division between star-forming and passive galaxies in  Appendix \ref{append:passive}.

We estimate the stellar mass completeness limit of our sample to be $\log_{10} (M_\star/M_\odot) = 9.7$, and we describe our methodology in the Appendix \ref{append:mass_completeness}. We show the mass limit with the vertical green dotted line in Figure \ref{fig:ms_passive_cut}.

We use existing SDSS classifications based on optical emission-line ratios to identify AGN, both for the \lcs \ and the field comparison samples. Specifically, we exploit the DR10 catalog\footnote{\url{http://www.sdss3.org/dr10/spectro/catalogs.php}} and flag as AGN the galaxies with SUBCLASS = AGN. This set is based on whether the galaxy has detectable emission lines that are consistent with being a Seyfert or LINER: $\rm \log(OIII/H\beta) > 0.7 - 1.2(\log(NII/H\alpha) + 0.4)$ \citep{Kauffmann2003}. 
We leave in galaxies that do not have a match to the DR10 AGN catalog.  
We test for the presence of AGN in the remanining sample using a WISE color cut, $W1 - W2 > 0.8$, from \citet{Stern2012}.  We find that only 0.2\% of the 497 galaxies in the cluster core and infall sample meets this color criteria.  We do not have WISE colors for the full field sample, but even if the fraction of AGN is a factor of two higher in the field \citep[e.g.][]{Kauffmann2004}, the overall contamination will be $< 0.5\%$.  Therefore we do not expect that obscured AGN (and the erroneous SFRs that we would infer from their SEDs) will introduce a large contamination in our sample.  {We do not remove the WISE AGN from the cluster sample so that we treat the cluster and field samples consistently.}

Both the field and \lcs\ samples are matched to the \citet{Simard2011} catalog so that we can utilize their measurements of effective radius and bulge-to-total ratio. \citet{Simard2011} performed two-dimensional bulge-to-disk decomposition for galaxies with 14$\leq m_{petro,r,corr} \leq$18 in SDSS DR7 using the GIM2D software.  We utilize their $r$-band bulge-to-total ratios ($B/T$) as a measure of galaxy morphology, and we use the fits where the bulge is fixed to be a de Vaucouleur profile with \sers \ index of $n=4$. 
We also use \sers \ indices and ellipticities from their single-component fits in the g-band. 

Hubble-type morphologies are  taken from \citet{Huertas-Company2011}, as reported by \cite{Tempel2014}.  They provide an automated morphological classification of the SDSS DR7 spectroscopic sample into four types (E, S0, Sab, Scd) based on support vector machines. 
They associate a probability to each galaxy of being in the four morphological classes instead of assigning a single class, to better reproduce the transition between different morphological types.


\subsection{The Final Samples}
In all the environments we consider only galaxies with a match to both the GSWLC and \cite{Simard2011} catalogs, and above the stellar mass completeness limit and SFR cut. We also exclude AGN.  Finally, we limit the sample to galaxies with ellipticities $e < 0.75$ to avoid highly inclined systems; we find some evidence that the GSWLC SFRs are systematically low for these galaxies, possibly due to internal extinction that is effective even at 12\micron. 
Then, we assemble two different sets of samples that will be used in our analysis. In Section \ref{sec:sfr_mstar_rel} we will consider all star-forming galaxies, regardless of their $B/T$ ratio. This selection results in a final core sample of 
137 
star-forming galaxies,  
an infall sample  of 360 
star-forming galaxies, 
and a  field sample with 11118 
star-forming galaxies.  In Section \ref{sec:disk_restrict}, we will consider only galaxies with $B/T \le 0.3$. 
In this case, the final core, infall, and field samples  include 86, 
216, 
and 7590 
star-forming galaxies, respectively.


\subsubsection{Mass Matching}
The mass distributions of the core and field samples differ at the $2.5\sigma$ level according to an Anderson-Darling test.  To ensure that our results are not driven by this marginal difference in the mass distributions, we compare the core sample with mass-matched field sample. To assemble the matched samples, we randomly select 30 galaxies from the field sample that fall in the range $\Delta \log_{10}(M_\star/M_\odot) < 0.15$ from each core galaxy. The mass offset of 0.15~dex is similar to the 0.13~dex uncertainly associated with the GSWLC-2 stellar masses (see \S\ref{sec:galaxy_properties}).  We draw 30 field galaxies for each cluster galaxy to retain the statistical advantage of the large field sample. The factor of 30 comes from the ratio of the field and infall sample sizes; the ratio of the field/core samples is $>70$, but we adopt the smaller ratio associated with the field/infall samples to be conservative and to simplify the process by adopting the same factor for all field/cluster comparisons. 
We draw randomly from the full field sample each time we select a galaxy, so our comparison field sample could include a given field galaxy multiple times. 
We repeat this process to create a field sample that is mass-matched to the infall sample.
Hereafter, we will always compare the cluster/field galaxies to mass-matched field samples.
Note, we do {\em not} create mass-matched samples of the core and infall galaxies because their mass distributions are not significantly different.


\section{Results}
\label{sec:results}
We present the distribution of SFRs and morphological parameters for the galaxies in our sample.  In Section~\ref{sec:sfr_mstar_rel}, we consider the samples of galaxies with no cut in $B/T$. In Section~\ref{sec:prop_suppressed} we discuss the $B/T$ distribution of galaxies with normal and suppressed star formation.
Finally, in Section~\ref{sec:disk_restrict}, we consider how the SFR distributions in different environments compare if we restrict our sample to disk-dominated galaxies only.  In Section~\ref{sec:disk-suppressed-prop} we discuss the properties of suppressed disk-dominated galaxies,  in Section~\ref{sec:phase-space-disks} we discussion how the SFRs depend on the phase-space location of galaxies in the cluster, and in Section~\ref{sec:altdiskID} we show how our results depend on our definition of disk-dominated galaxies.  Our separate treatment of disk-dominated galaxies is a critical part of our analysis as it demonstrates that the observed suppression of SFRs in cluster galaxies is not simply reflecting a correlation of morphology with environment.

\begin{figure*}
    \centering
    \includegraphics[width=.45\textwidth]{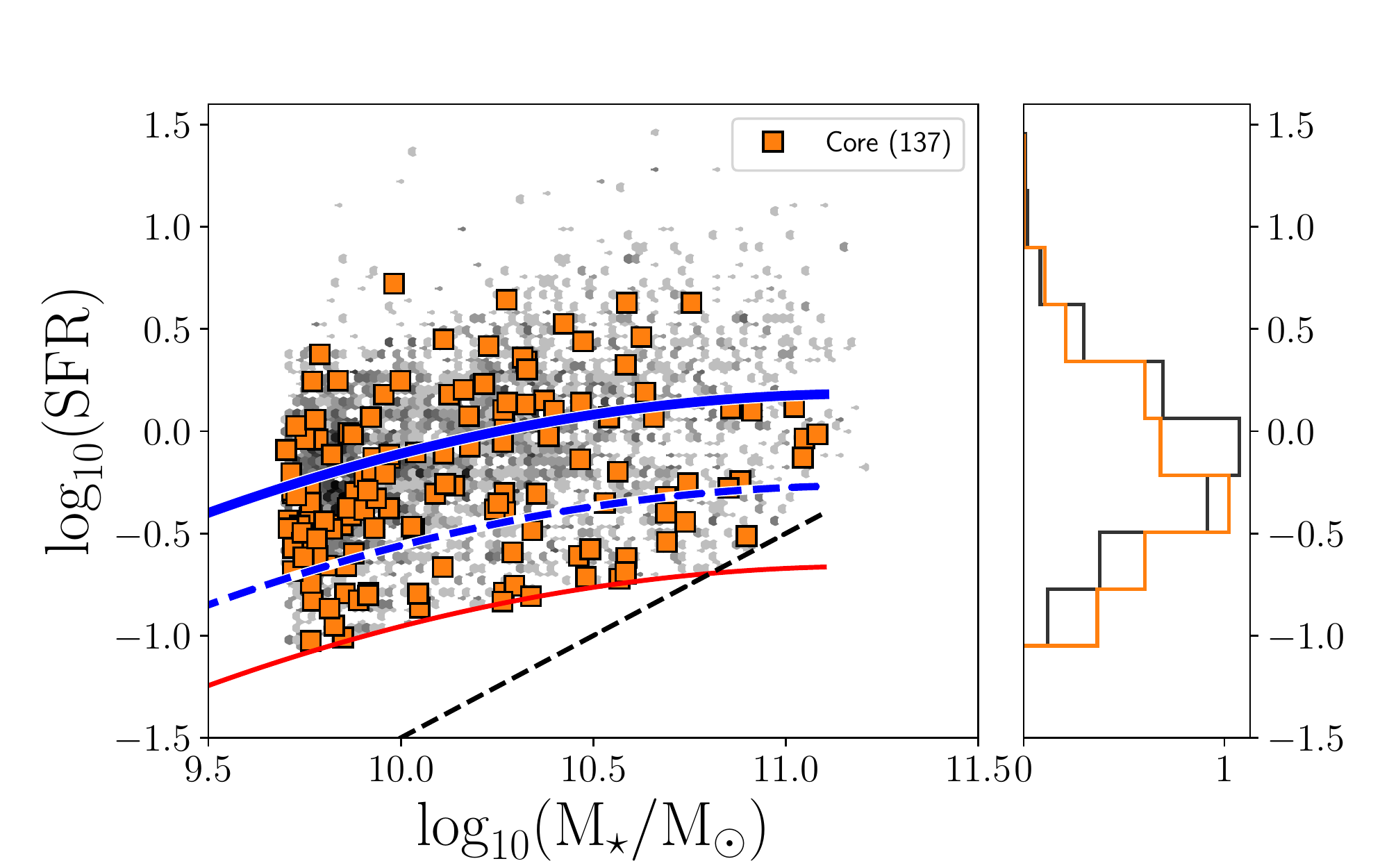}
    \includegraphics[width=.45\textwidth]{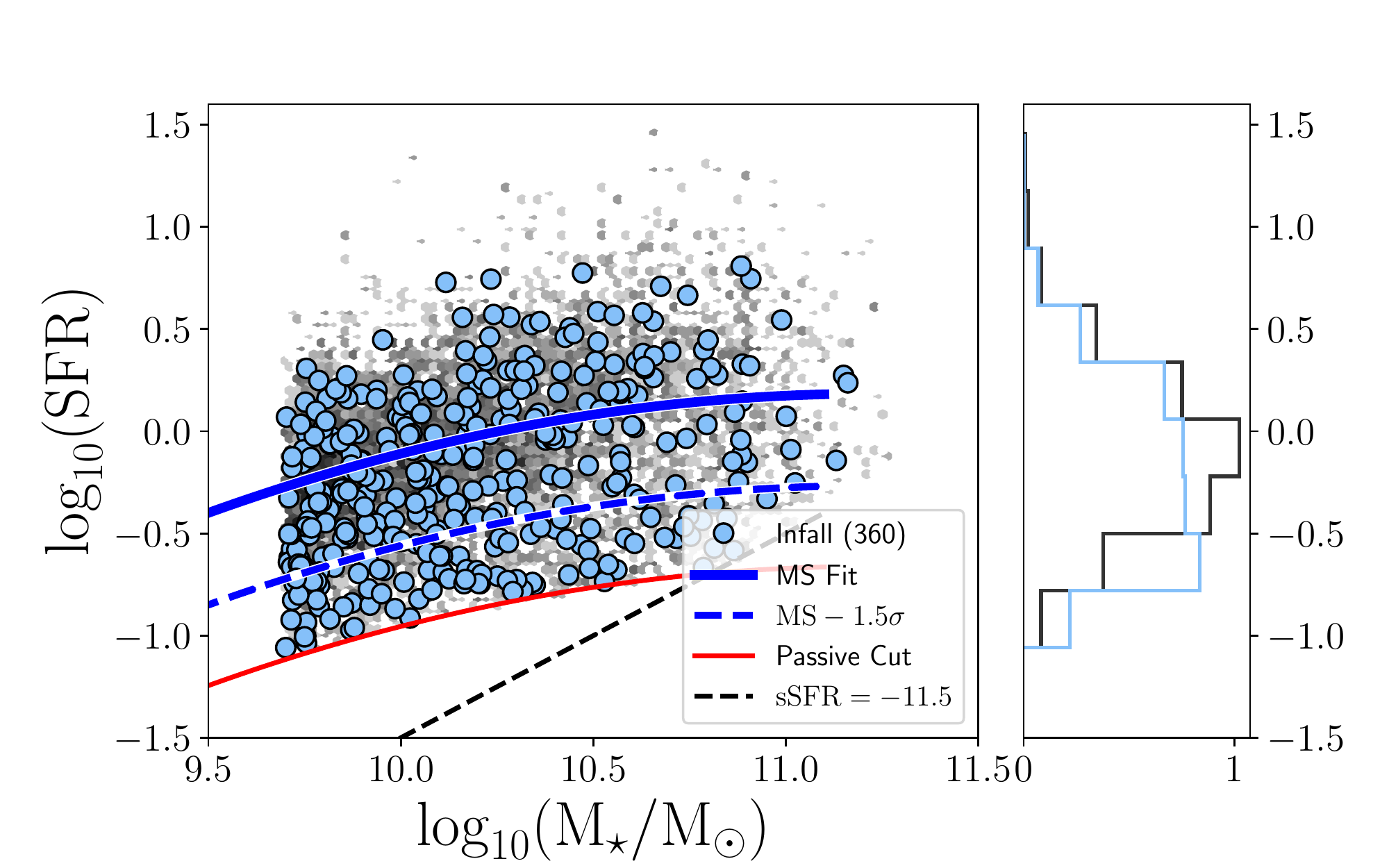}
    \caption{(Left) $\log_{10} (SFR)$  versus $\log_{10}(M_\star/M_\odot)$ for LCS core (orange) and field galaxies (gray).  
    The solid blue line shows our fit to the $SFR-M_\star$ relation, and the dashed blue line shows the fit minus $1.5\sigma$ (see Appendix \ref{append:sfms_fit} for details).  The red line shows the division we adopt between star-forming and passive galaxies (see Appendix \ref{append:passive} for details), and the black dotted line is the specific SFR limit of the GSWLC catalog ($sSFR = -11.5$).  The 
    core galaxies have a statistically significant excess population of low SFR galaxies compared with the field.   
       (Right) Same as left panel but for 
       infall (light blue) and field galaxies (gray).  Again, the 
       infall galaxies have a significantly larger fraction of galaxies with low SFR compared with the field. 
    }
    \label{fig:lcsgsw-sfrmstar}
\end{figure*}

\subsection{SFR-Stellar Mass Relation of Star-Forming Galaxies in Different Environments}
\label{sec:sfr_mstar_rel}

\input{table1-massmatch}
A way to understand 
the physical processes at the origin of quenching is to investigate how the SFRs of star-forming galaxies of a given stellar mass vary in different environments. 
Figure \ref{fig:lcsgsw-sfrmstar} shows the SFR-mass relation of galaxies in the cluster cores, infall, and field regions.  
The main plot of the left panel 
presents the results for  
the core (orange squares) and field samples (gray points), while the right inset shows the normalized distributions of SFR and stellar mass.
The SFR histograms show that 
while the core galaxies can have SFRs as high as the field galaxies, the core sample includes a population of galaxies with reduced SFRs that are less prevalent in the field.  
We compare the SFR distributions of the core and field samples using an Anderson-Darling test 
and find with high confidence ($>19\sigma$) that they are drawn from different parent samples.


Next, we compare the infall galaxies with the field sample in the right panel of Figure \ref{fig:lcsgsw-sfrmstar}.   Again, the SFRs of the field and infall samples differ significantly ($>30\sigma$ according to Anderson-Darling test). 
Finally, we compare the core and infall samples in terms of both the SFR and stellar mass (not shown).  
We are not able to distinguish the SFR distributions. 
In summary, the SFRs of the both the core and infall galaxies differ from the field, but not from each other. 
The statistics are reported in Table \ref{tab:stats} (center column).  

\begin{figure}
    \centering
    \includegraphics[width=.45\textwidth]{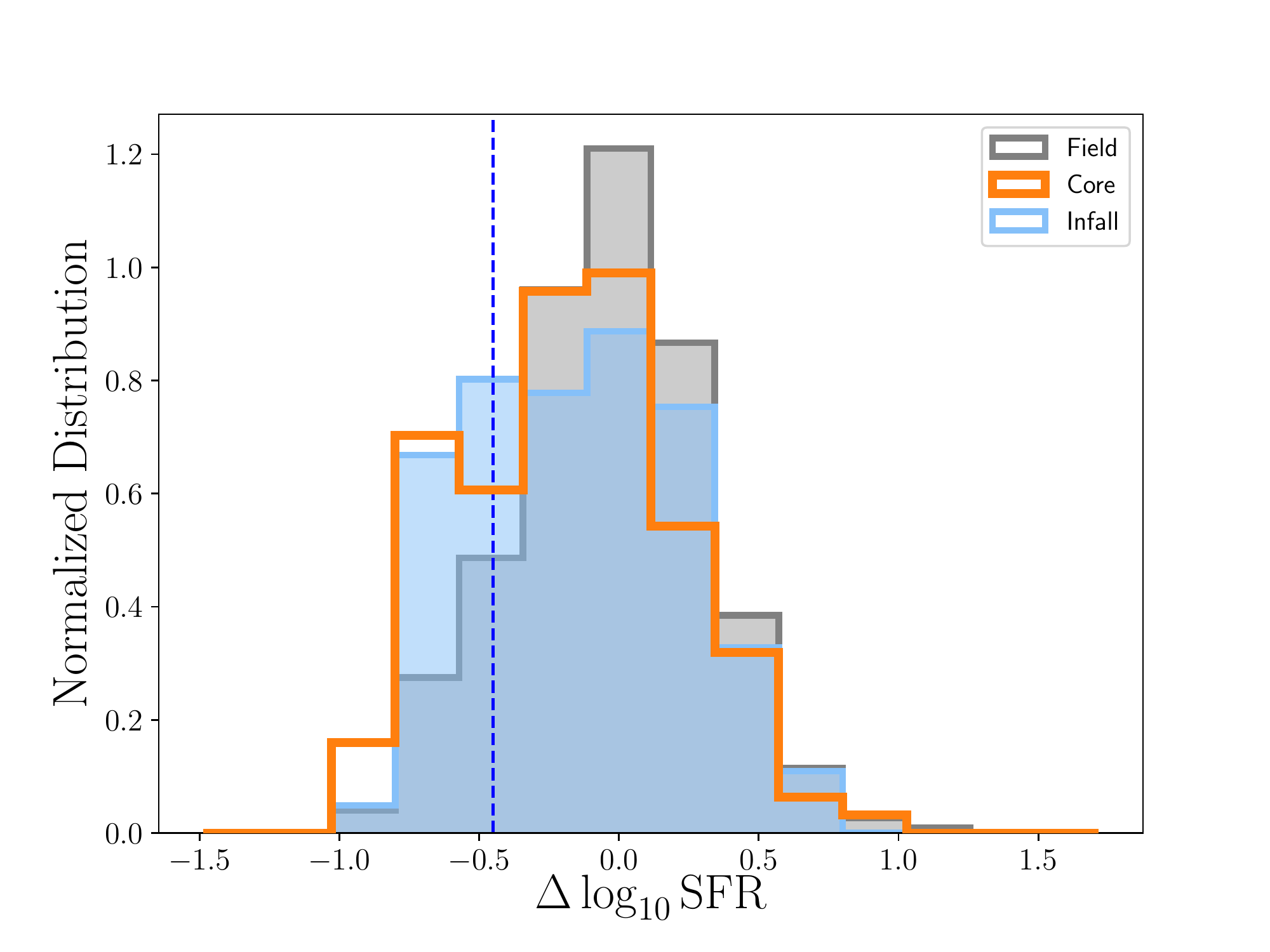}
    \caption{Normalized histogram of \dsfr \ (Eqn. \ref{eqn:dsfr}) for the core (orange), infall (blue), and mass-matched field (gray) samples. (The histograms are normalized so that the integral is one.) Galaxies are included regardless of their $B/T$.
      The core and infall samples show an excess of galaxies with suppressed SFRs relative to the field, and an Anderson-Darling test confirms this difference (Tab.~\ref{tab:stats}).  The blue, vertical dashed line shows $- 1.5\sigma=-0.45$~dex below the main sequence fit.  We consider galaxies below $\Delta \log_{10}(SFR) < -0.45$ to be suppressed (Sec. \ref{sec:sfr_mstar_rel}). 
    }
    \label{fig:dsfr-hist}
\end{figure}

We further examine the distribution of SFRs at a fixed stellar mass by comparing the offset of each galaxy's SFR with respect to the star-forming main sequence:  
\begin{equation}
\Delta\log_{10} SFR = \log_{10}(SFR_{obs}) - \log_{10}(SFR_{MS}(M_\star)),
\label{eqn:dsfr}
\end{equation}
where $\log_{10}(SFR_{MS}(M_\star))$ is the value predicted from the fit based on its stellar mass (Eqn. \ref{eqn:sfms}).
The normalized distribution of such differences is shown in Figure \ref{fig:dsfr-hist} for the core, infall, and mass-matched field samples.  Here the field is mass-matched to the combined infall and core sample.   
Both the core and infall samples show an excess of galaxies with suppressed SFRs relative to the field. 
The mean values of $\Delta \log_{10}(SFR)$ for the field, infall, and core samples are: $-0.040\pm0.003$, 
$-0.17\pm0.02$, and 
$-0.20\pm0.03$, where the error is the standard error in the mean. 
We find similar results when comparing the median, although the errors as estimated from bootstrap resampling are larger (median $\Delta \log_{10}(SFR)$ for the field, infall, and core samples: 
$0.031^{+0.005}_{-0.006}$, $-0.17^{+0.05}_{-0.06}$, and $-0.21^{+0.07}_{-0.02}$, where the error is the 68\% confidence interval from bootstrap resampling).  
An Anderson-Darling test shows that the core and infall samples are significantly different from the field but not from each other (p values reported in Table~\ref{tab:stats}). 
 
\begin{figure}
   \centering
   
   \includegraphics[width=.45\textwidth]{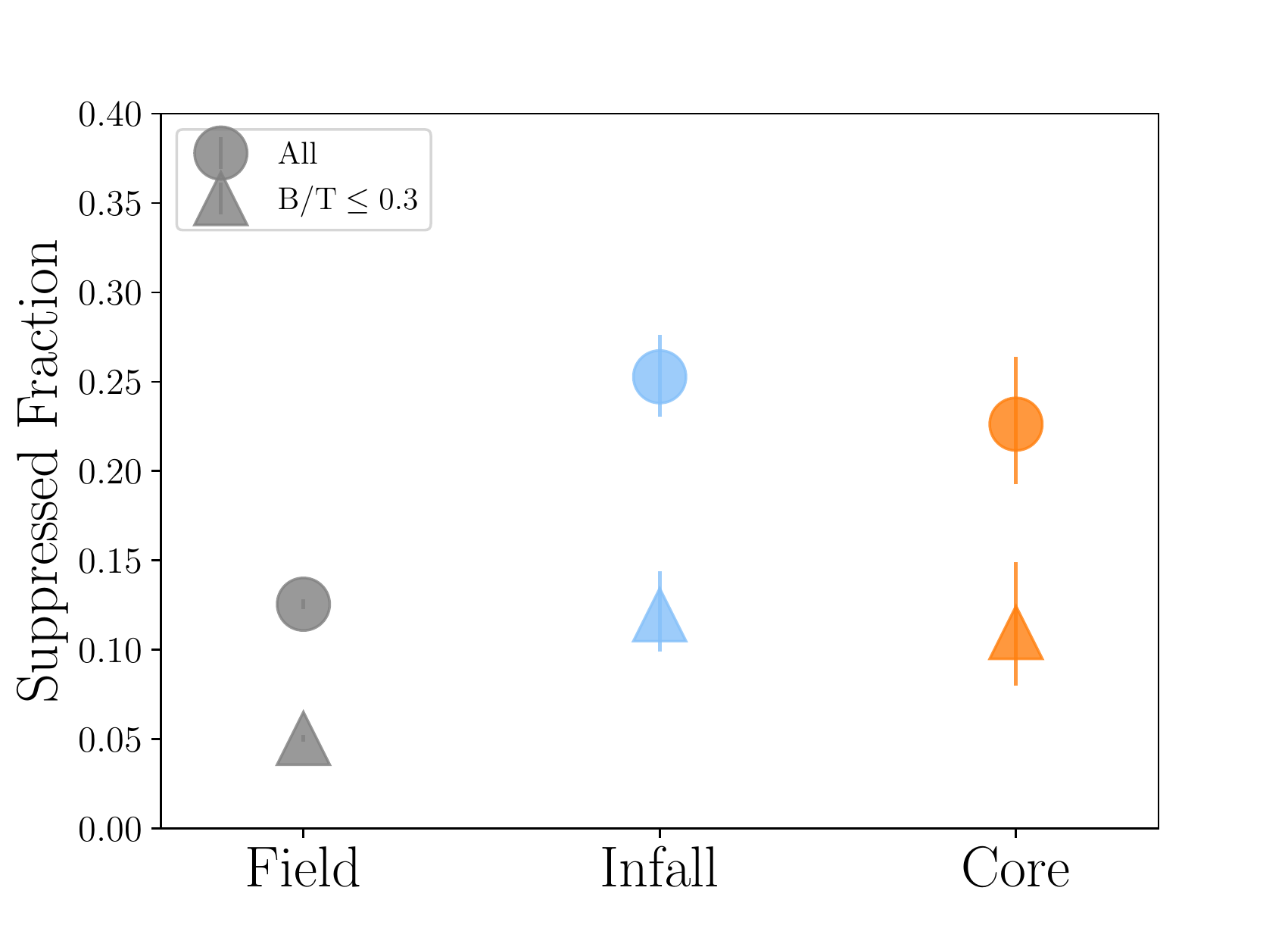}
    \caption{The fraction of star-forming galaxies with suppressed star-formation vs. environment.  The circles and triangles are for the full and $B/T \le 0.3$ (\S\ref{sec:disk_restrict}) samples, respectively.  The fraction of suppressed galaxies increase from the field to the cluster.
}
    \label{fig:fsuppressed}
\end{figure}

Following \citet{Paccagnella2016}, we use the offset from the fit to the field main sequence to identify galaxies with suppressed SFRs, associating those with galaxies that are likely transitioning to the passive population. We define {\it normal} star-forming galaxies as those with SFRs within $1.5\sigma$ of the best-fit SFR-$\rm M_\star$ relation, and we measure the standard deviation to be $\sigma = 0.3$.  We define {\it suppressed} galaxies as those with SFRs that fall 1.5$\sigma$ below the relation ($\rm \Delta \log SFR < -0.45$~dex). We show this division with the dashed blue line in Figures \ref{fig:lcsgsw-sfrmstar} and \ref{fig:dsfr-hist}. 
According to this definition, we would expect to find 7\% of the population with suppressed SFRs if SFRs are normally distributed about the best-fit relation.  Instead, we find a higher fraction of suppressed galaxies in all environments, and the fraction is higher in the core and infall regions than the field, as shown in Figure \ref{fig:fsuppressed}.  The fraction of suppressed galaxies is 13.1$\pm$0.3\% in the field, 27$\pm 2$\% in the infall region, and  25$\pm 4$\% in the cluster.  It is interesting that the fraction of suppressed galaxies increases already in the infall region with with no additional increase when moving to the core.  With our limited sample size we are not able to distinguish if this reflects a true lack of change in the fraction of suppressed galaxies or a change in the distribution of SFRs or if it results from poor statistics.  In Section~\ref{sec:discussion} we discuss how the decline in SFR in different environments may be related to the infall time. {In Section~\ref{sec:passive-contamination} we discuss how our ability to measure the fraction of suppressed galaxies could be affected by contamination from passive galaxies.  }

\subsection{B/T distribution of suppressed galaxies}
\label{sec:prop_suppressed}

\begin{figure}
    \centering
    \includegraphics[width=.5\textwidth]{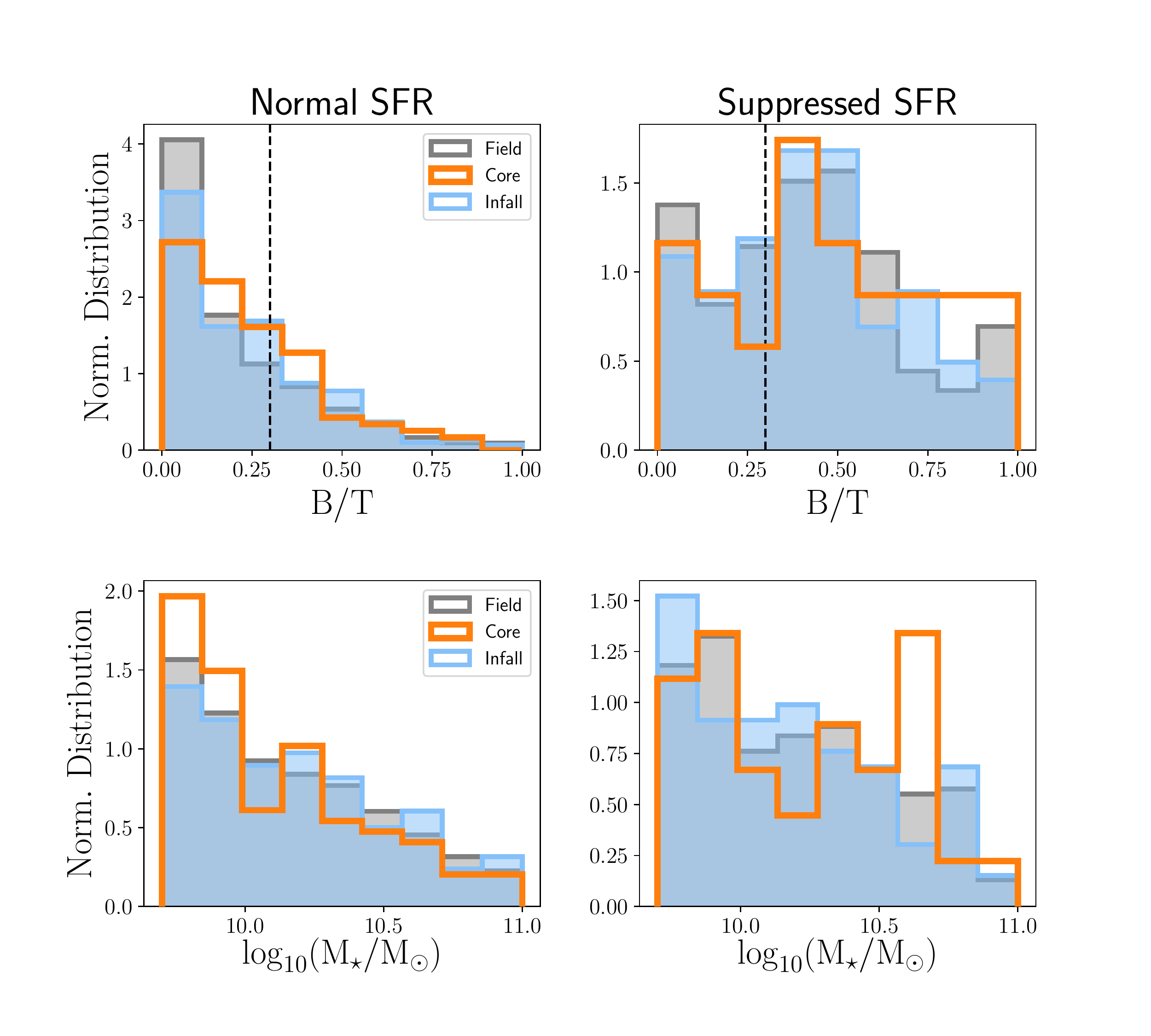}
    \caption{$B/T$ distribution (top) and mass distribution (bottom)  of normal (left) and suppressed-sfr (right) galaxies, with field, infall, and core galaxies shown with gray, blue and orange histograms. For all four panels in this figure we have mass-matched the total field sample with the combined core and infall samples but have not mass-matched normal and suppressed galaxies.  In all environments, the distribution of $B/T$ values for suppressed galaxies is shifted toward higher values relative to normal galaxies. The mass distributions of the normal and suppressed core/infall galaxies are statistically consistent with each other, but the mass distributions of the normal and suppressed field galaxies are significantly different.
    }
    \label{fig:prop_lowsfr_field_cluster}
\end{figure}

 In the previous subsection we have shown that the fraction of suppressed galaxies depends on environment; in this section, we investigate the morphological properties of the suppressed and normal galaxies in each environment. 
 In Figure \ref{fig:prop_lowsfr_field_cluster} we compare the $B/T$ (top)  and stellar mass (bottom) distributions of  normal (left) and suppressed  (right) galaxies, showing the distribution in each environment.  Note that while the field is mass-matched to the combined core and infall samples, we do not match based on normal vs. suppressed SFRs.  The top row shows the most {significant result of this figure} - that in all environments, galaxies with suppressed SFRs have higher $B/T$ ratios than galaxies with normal SFRs.  For all three samples, an Anderson Darling test confirms that the SFRs of normal vs suppressed galaxies are significantly different, with p-values $<0.001$.
We show the mass distributions of normal vs. suppressed galaxies in the bottom row in Figure \ref{fig:prop_lowsfr_field_cluster}.  An Anderson-Darling test shows that the mass distributions of the normal and suppressed infall/core samples are indistinguishable, while those of normal and suppressed field galaxies are significantly different (p value $<0.001$ according the Anderson-Darling test). The stellar mass distribution of normal SFR field galaxies is skewed towards low-mass galaxies compared to that of suppressed galaxies. 
 Therefore, we have no evidence to support the hypothesis that the observed difference in the $B/T$ distributions of the normal and suppressed core and infall galaxies are driven by differences in stellar mass distributions.  To confirm that this is also the case for the field, we
create a subset of the normal field galaxies that is mass-matched to the suppressed field sample, using the same mass-matching procedure as described in Section \ref{sec:sfr_mstar_rel}. We find that the mass-matched suppressed field galaxies have significantly higher $B/T$ values, showing that suppressed field galaxies are more likely to have higher bulge fractions than normal field galaxies of comparable stellar mass.

\begin{figure}
    \centering
    \includegraphics[width=0.5\textwidth]{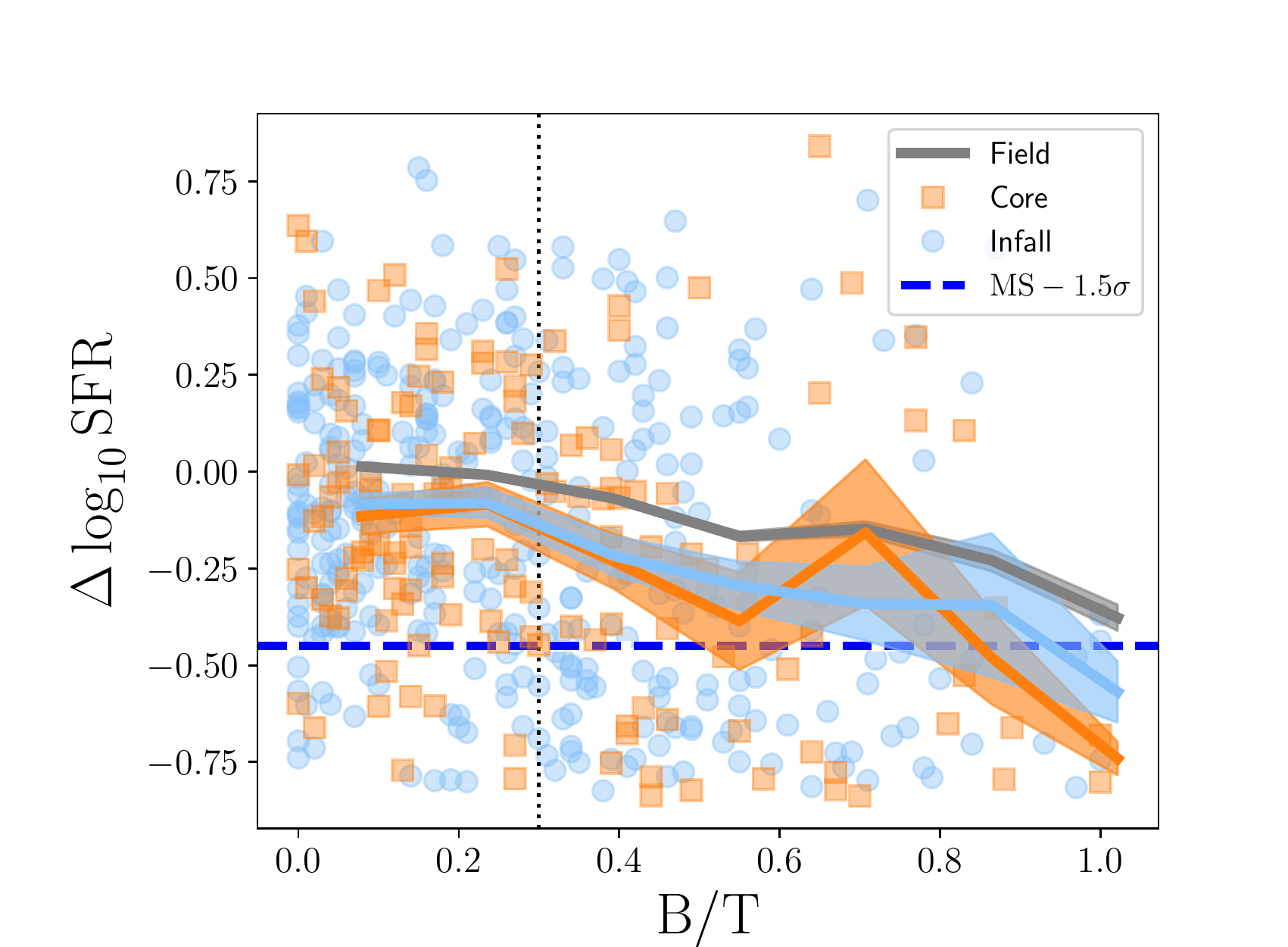}
    \caption{\dsfr \ versus $B/T$ for core (orange square), infall (light blue circles), and mass-matched field samples (gray).  Core and infall SFRs are systematically below the field, at all $B/T$. The dashed line shows the adopted separation between normal and suppressed galaxies. The lines show the binned mean, and the shaded regions show the error  in the mean. 
    }
    \label{fig:dsfr-bt}
\end{figure}

These results, 
in combination with the well-established morphology-density relation \citep[e.g.][]{Dressler1980,Postman2005,Fasano2015,Vulcani2023}, could suggest that the difference in the SFR distributions vs. environments that we established in Section \ref{sec:sfr_mstar_rel} is primarily due to differences in the $B/T$ distributions. To illustrate this point, Figure \ref{fig:dsfr-bt} shows $\Delta {\rm log_{10}}\rm  SFR$ versus $B/T$ for the core, infall, and mass-matched field. Indeed, there is a strong relation between these two parameters, with higher $B/T$ galaxies having lower \dsfr.  Nonetheless, at a fixed $B/T$ the \dsfr \ of the core and infall galaxies are systematically lower than those of the field galaxies.  We will discuss the implications of this further in \S\ref{sec:discussion}.

\subsection{Restricting to Disk-Dominated Galaxies}
\label{sec:disk_restrict}

While Figure~\ref{fig:dsfr-bt} clearly shows that \dsfr \ is reduced in dense environments with respect to the field at all values of $B/T$, the joint dependence of SFR on both morphology and environment complicates any further discussion of what physical processes are at play. To reduce the effect of this joint dependence, we therefore restrict our analysis to galaxies with  $B/T \le 0.3$.  
This $B/T$ cut corresponds to the point in Figure~\ref{fig:dsfr-bt} below which the $\Delta {\rm log_{10}} SFR$-$B/T$ relation is rather flat.  For $B/T > 0.3$, $\Delta {\rm log_{10}} SFR$ drops to lower SFRs.  
We repeat the analysis performed in Sec.\ref{sec:sfr_mstar_rel} to see how SFR varies with environment for disk-dominated galaxies only. 
Regarding the $SFR-M_\star$ relation, the difference in the SFR distributions between the core/infall and field samples persists (plots shown in Appendix \ref{append:sfms_fit}).  The supporting statistics are summarized in the last  column of Table \ref{tab:stats}.

\begin{figure}
    \centering
    \includegraphics[width=.45\textwidth]{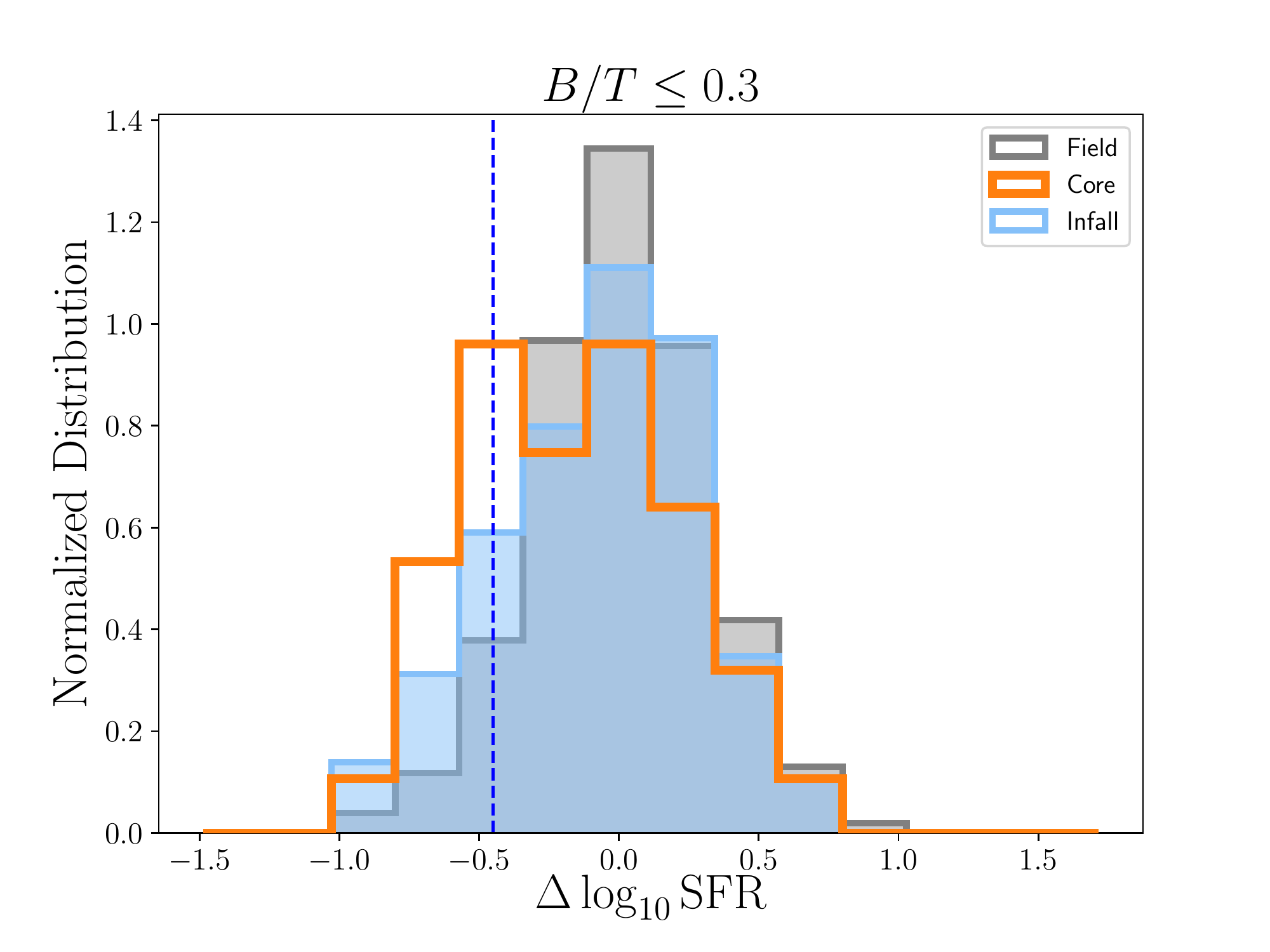}
    \caption{Same as Fig. \ref{fig:dsfr-hist} 
    but only for  $B/T \le 0.3$ galaxies. 
    Even after restricting to disk-dominated galaxies, the core and infall samples shows an excess of galaxies with suppressed SFRs relative to the field (Tab.~\ref{tab:stats}).  The blue, vertical dashed line shows $\- 1.5\sigma=-0.45$~dex below the main sequence fit.
    \label{fig:dsfr-hist-btcut} }
\end{figure}

We refit the SFR-Mass relation using only the $B/T \le 0.3$ field galaxies as discussed in the Appendix \ref{append:sfms_fit}.  
We recalculate \dsfr \ relative to the new main sequence fit, and we show the resulting distributions in Figure \ref{fig:dsfr-hist-btcut}.  The rms for the disk-only main sequence is 0.28~dex, which is slightly smaller than for the full sample. As with the full sample, both the core and infall galaxies present a tail in the \dsfr \ distribution that is not seen in the mass-matched field. The differences between the core vs. field and infall vs. field are statistically significant according to an Anderson-Darling test (Table \ref{tab:stats}).  We are not able to distinguish the infall and field samples.
Disk dominated suppressed galaxies 
represent 11$\pm 4$\% of the star-forming cluster core population, 12$\pm2$\% of the infall sample, and only 5.1$\pm$0.2\% of the field population (Fig. \ref{fig:fsuppressed}). 
A dependence on environment is still evident as the fraction of suppressed galaxies is significantly elevated in both the core and infall regions relative to the field. 

\subsubsection{Properties of Suppressed Disk-Dominated Galaxies}
\label{sec:disk-suppressed-prop}
\begin{figure*}
    \centering
     \includegraphics[trim = 0 50 0 10, width=0.95\textwidth]{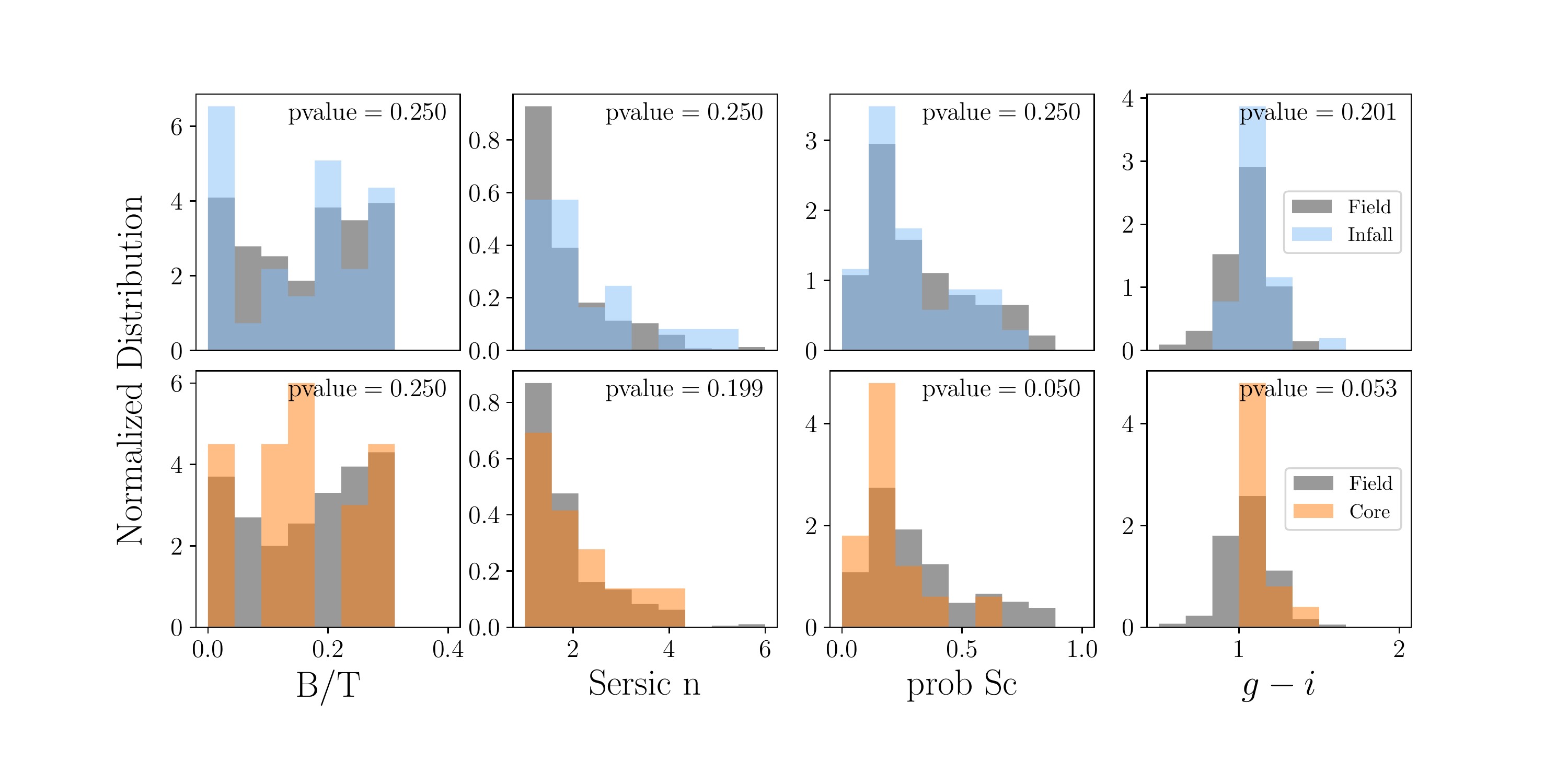}
    \caption{(Top) Comparison of suppressed $B/T \le 0.3$ infall galaxies (blue) with a subsample of suppressed field galaxies (gray) that are mass-matched to the infall sample. The p value from the Anderson-Darling test is shown at the top of each panel.
    The infall and mass-matched field samples are indistinguishable in terms of $B/T$, \sers \ index, the probability of being classified as an Sc galaxy, and $g-i$ colors.
    (Bottom) Same as above, but comparing suppressed {\em core} galaxies (orange) with a mass-matched sample of the low-SFR field galaxies.  There is a hint that the core galaxies have a lower probability of being classified as an Sc galaxy and redder $g-i$ colors, but the differences are not statistically significant. Morphology and color of suppressed galaxies appears to be largely independent of environment.  
    }
    \label{fig:properties_lowsfr_gals}
\end{figure*}

We  now
compare the properties of the suppressed galaxies as a function of environment.  Specifically, we examine other measures of morphology in addition to $B/T$ like \sers \ index and the probability of being an Sc galaxy, to see if the suppressed cluster galaxies differ structurally from the suppressed field galaxies.  In addition, we compare optical $g-i$ colors to look for global differences in stellar populations.
We are only comparing the disk-dominated galaxies, so we do not expect a large variation in morphological properties with environment.  However, the colors of the disk-dominated core galaxies could be different \citep[e.g.][]{Weinmann2009, Cantale2016}.
To perform comparisons not driven by the different mass distributions of the samples, we first 
select a subsample of the suppressed field sample with the same stellar mass distribution as the suppressed infall galaxies.  
We compare the properties of these two populations in the top row of Figure \ref{fig:properties_lowsfr_gals} and find that the suppressed infall galaxies (blue) and mass-matched field sample (gray) are {statistically consistent} in terms of $B/T$, \sers \ index, the probability of being an Sc galaxy, and $g-i$ color.  {We give the Anderson-Darling p-values in each panel in \ref{fig:properties_lowsfr_gals}, demonstrating our inability to reject the null hypothesis that the distributions in the field and infall regions are identical.}
We repeat the mass-matching process between the suppressed field and core samples
and show the results in the bottom panel of Figure \ref{fig:properties_lowsfr_gals}.  
There is a hint that the core galaxies have a lower probability being classified as an Sc galaxy and have redder $g-i$ colors than the suppressed field galaxies, but these differences are not statistically significant. 
In summary, the morphological properties of the suppressed, disk-dominated galaxies are not a strong function of environment.  

\subsubsection{Phase-Space Distribution of Low-SFR Disk Galaxies}
\label{sec:phase-space-disks}
\begin{figure*}
    \centering
    \includegraphics[width=\textwidth]{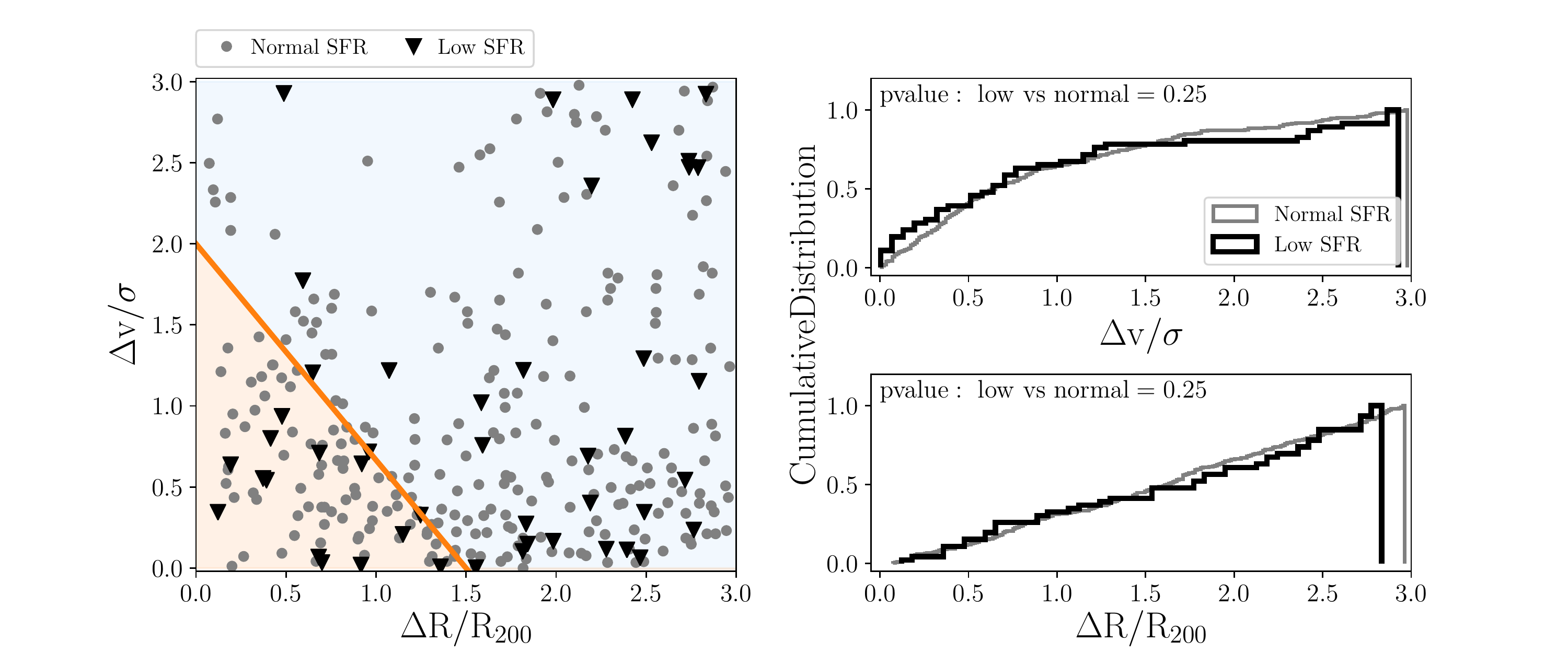}
    \caption{(Left) $\Delta v/\sigma$ vs. $\Delta R/R_{200}$ for $B/T \le 0.3$ core and infall galaxies with normal (gray circles) and suppressed SFRs (black triangles).  The diagonal orange line shows the division from \citet{Oman2013} that we use to separate core and infall galaxies.   (Upper Right)  Cumulative distribution of $\Delta v/\sigma$    and (Lower Right) $\Delta R/R_{200}$ for galaxies with normal (gray) and suppressed SFRs (black). The normal and suppressed galaxies are indistinguishable in terms of both $\Delta R/R_{200}$  and $\Delta v/\sigma$ (the p value from the Anderson-Darling test is reported in each panel).  
    \label{fig:phasespace}}
\end{figure*}

The position of a galaxy in a phase-space diagram is strongly correlated with the time that the galaxy has been in the cluster environment, such that galaxies that have been in the cluster environment for the longest time are preferentially found at low values of both projected cluster-centric radii and velocity offset from the mean cluster velocity \citep[e.g.][]{Oman2013,Jaffe2015}.  As discussed in Section \ref{sec:sample}, we use this information to define our core and infall samples.  We now revisit this characterization to see if the $B/T \le 0.3$ galaxies with suppressed SFRs occupy a distinct region in the phase-space diagram.
In the left panel of Figure \ref{fig:phasespace}, we compare the location of the galaxies with normal (gray circle) and suppressed SFRs (black diamond) in phase space.  We show the cumulative histograms of $\Delta R/R_{200}$ and $\Delta v/\sigma$ in the right two panels. The distributions of normal and suppressed star-forming galaxies are indistinguishable in terms of both $\Delta v/\sigma$ and $\Delta r/R_{200}$, respectively.  The results are unchanged if we consider the full sample with no restriction on $B/T$.

\subsubsection{Alternate Methods to Identify Disk Galaxies}
\label{sec:altdiskID}
Previous authors have identified limitations with using bulge-disk decomposition to isolate disk galaxies.  For example, \citet{Meert15} and \citet{Cook19} point out that off-centered bulges and disks, strong secondary features, bars, or isophotal twists can all result in apparent photometric bulges that are not linked to physically distinct structures.  This would result in true disk galaxies being falsely classified as bulge-dominated galaxies.  We  explore this issue using the \citet{Sersic63} index, $n$, as an alternate way of selecting disk-dominated galaxies.  This is a common method of defining disk-dominate galaxies based on their light profiles.  Common values for a cut range from $n=1-3$ \citep{Allen06,Kelvin12}.  We adopt a cut set at $n<2.5$ following \citet{Cook19}.   The main-sequence fit for the $n<2.5$ cut is indistinguishable from our default value based in $B/T$.  With the $n<2.5$ cut, the observed differences in ${\rm log_{10}} SFR$ and $\Delta {\rm log_{10}} SFR$ persist with high ($>3\sigma$) significance. 

A different issue is related to the relative fading of bulges and disks.  If a young disk has its star formation rapidly truncated, it will fade more quickly than an old bulge, thus increasing the apparent $B/T$ from a pure mass-to-light effect and without any redistribution of the stellar mass.  This could in principle move galaxies beyond our $B/T=0.3$ cut as they quench.  This scenario was investigated by \citet{Christlein04}, who found that early type galaxies in clusters, including S0 galaxies, could not be formed by fading disks but rather had to experience a growth in the bulge light.  While this is not an obvious outcome of the quenching process, it could occur if gas is driven to the center by ram-pressure stripping and if the stars are subsequently perturbed from circular orbits by galaxy harassment.  On the other hand, a more recent study by \citet{Vulcani15} showed that quenching in group galaxies is not accompanied by a change on morphological structure. It is not straightforward to model how spatially dependent quenching could affect our results.  We therefore make the assumption, supported by \citet{Vulcani15}, that  the quenching process does not affect galaxy structure enough to alter our result.  {We note however that there is a significant amount of disagreement in the literature as to the degree to which morphological evolution is accompanied by the quenching of star formation \citep[e.g.][]{Cappellari13,Cortese19,Croom21,Park22}.}

\section{Quantifying the Timescale of SFR Decline} \label{sec:model}
In the previous sections, we found evidence for an excess population of galaxies with suppressed SFRs in the infall and core regions with respect to the field.
In this section, we model the evolution of the integrated SFRs to constrain  the timescale associated with the physical processes that could be driving this suppression in SFRs. We completed a similar exercise in \citet{Finn2018} by comparing the size of the star-forming disks as traced by 24\micron \ sizes;  in this paper we use integrated SFRs to provide independent constraints.  
\subsection{Modeling Cluster Infall}

We characterize the decline in SFRs of the cluster core galaxies using a model that is based
on the best-fit model from \citet{Wetzel2013}.
In the \citet{Wetzel2013} model, galaxies fall into the cluster and remain unaffected for some specified delay phase period.  The galaxies then experience a rapid quenching event.
In our model, we create simulated core galaxies using the general framework from \citet{Wetzel2013}.  During the delay phase, the SFRs of the infalling galaxies follow the redshift-evolution of field SFRs.  We model the quenching timescale with an exponential decay rate of $\tau$, and we determine acceptable values of $\tau$ at each value of the delay phase.  A combination of delay time plus $\tau$ is acceptable if an Anderson Darling test is not able to distinguish the SFR distribution of the observed core galaxies from the simulated core galaxies.

To quantify the delay and quenching times, we first need to adopt a timescale for how long the star-forming core galaxies have been in the cluster environment.  
According to \citet{Oman2013}, galaxies in the region of phase-space that we use to define the core sample are likely to have been within $2.5$ times the cluster virial radius for $3-7$~Gyr.\footnote{We do not know if the time since infall in the core region of phase-space holds for star-forming galaxies.  Determining this time from simulations would be highly model dependent as it would rest on the ability of the simulations to accurately model the SFHs of infalling galaxies.  We prefer to avoid such a strong reliance on the accuracy of cluster-galaxy SFHs in simulations, something which has not be well-tested.  With this caveat in mind, we assume that star forming galaxies at low cluster-centric radius and velocity offset have statistically spent the longest time in the cluster environment.  We account for potentially shorter time-since infall for star-forming galaxies with the large range in infall times allowed by our modeling.} Thus, we adopt $7$~Gyr as the maximum time since infall, and we refer to this parameter as $t_{max}$.  Accretion rates are fairly uniform over this timescale \citep{McGee2009}, and this allows us to assign our simulated galaxies infall times that range uniformly from zero to $t_{max}$.  We also assume that the accretion timescale is independent of stellar mass.
We let the delay time, $t_{delay}$, vary between zero and $t_{max}$, and we use the same delay time for all galaxies in a particular step.
We refer to the time period during which quenching occurs as $t_{active}$, where
\begin{equation}
  t_{active} = t_{infall} - t_{delay}.
\end{equation}
If $t_{delay} > t_{infall}$ for a particular galaxy, then we set $t_{active} = 0$.

To determine the SFR and stellar mass distributions of the simulated core sample, we start with the $z=0$ field sample.
We evolve the $z=0$ field galaxies back in time, adjusting their SFR and stellar masses to what they would have been at the time when quenching begins (at $t_{active}$).  To determine the change in SFR and stellar mass, we first forward model a grid of galaxies with a range of stellar masses and sSFRs from 7~Gyr ago to the present.  For each galaxy, we evolve SFRs according to the evolution of the field SFR-$M_{\ast}$ relation as parameterized by \citet{Whitaker2012}: 
\begin{equation}
\label{eqn:whitaker}
    \log_{10} SFR = \alpha(\log_{10} M_{\star} - 10.5) + \beta,
\end{equation}
where $\alpha = 0.70-0.13z$ and $\beta = 0.38 + 1.14z - 0.19z^2$.
We increment the stellar mass according to the time-evolving SFR.  We also account for stellar mass loss at each time step as described in \citet{Poggianti2013}, where the fraction of stellar mass retained by a simple stellar population is
\begin{equation}
    M_{retained} = 1.749 - 0.124~\log_{10}(t_{yr})
\end{equation}
for stellar populations with ages greater than $1.9\times 10^6$~yr.
With the grid of galaxy stellar masses, SFRs, and their time evolution in place, we are able to link each field galaxy at $z=0$ with its progenitor at the time when quenching begins.  We refer to this as the pre-quenching sample.

We then evolve the pre-quenching sample forward to $z=0$, modeling the effect of the cluster environment on their SFRs
with an exponential function:
\begin{equation}
\label{eqn:cluster-sfr-decline}
    SFR_{sim-core} =  SFR_{infall-field}~ e^{-t_{active}/\tau} ,
\end{equation}
We step through e-folding times ranging from $0.5 <\tau <  7$~Gyr.
The stellar mass increases based on the integral of $\int _{t_{active}}^{0} SFR(t) ~dt$, and we again apply the \citet{Poggianti2013} prescription for stellar mass loss. We refer to the resulting $z=0$ sample as the simulated core (sim-core) sample.  

For each combination of $t_{delay}$ and $\tau$, we mass-match the simulated core sample to the actual core galaxies by randomly selecting 60 of the sim-core galaxies with a stellar mass offset less than 0.15 dex from each core galaxy.  We select 60 simulated galaxies rather than one to reproduce the statistical advantage of the large field sample that we are using to create the simulated core galaxies.   
We reproduce the effect of our observational SFR limits by removing any sim-core galaxies with SFRs below our SF/passive cut (below red line in Figure \ref{fig:ms_passive_cut}); in effect, these galaxies would be fully quenched by $z=0$ according to our definition.
We then compare the SFR distributions of the sim-core and core samples.  
We define an unacceptable model as one with an Anderson-Darling p-value less than $p < 0.003$; this indicates that the distribution of core and sim-core SFRs for that particular $t_{delay}-\tau$ pair are significantly different. 

\begin{figure}
    \centering
    \includegraphics[width=0.5\textwidth]{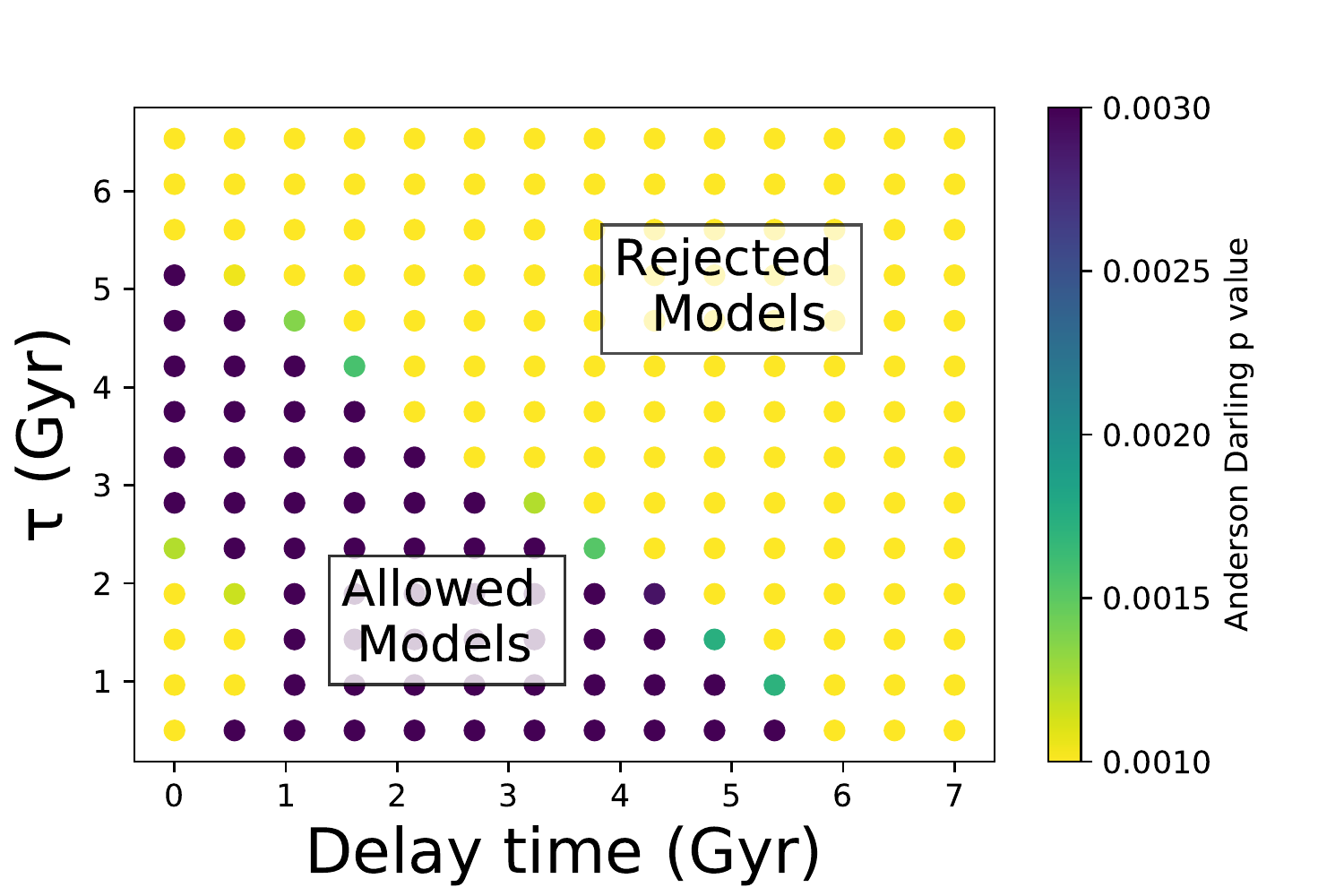}
    \caption{The exponential decay time associated with SFR decline in the cluster environment, $\tau$, versus the delay time.  The points are color-coded by the p value of the Anderson-Darling test from comparing the SFR distribution of the simulated and observed core populations. }  
    \label{fig:sfr_pvalue}
\end{figure}

In Figure \ref{fig:sfr_pvalue}, we show $\tau$ versus $t_{delay}$, with points color coded by their p value.  Each point corresponds to one model.  Purple points show acceptable models with $\rm p \ values > 0.003$, meaning the SFR distributions of the core and sim-core galaxies are not significantly different.  The yellow points show unacceptable models ($\rm p \ values < 0.003$).  The modeling shows several interesting results.  First, if there is no delay period, then the environmental quenching proceeds with a relatively long e-folding time between $3 < \tau < 5$~Gyr.  Second, as the delay time increases, the e-folding time associated with active quenching decreases; in effect, this is by construction because we require $t_{infall} = t_{delay} + t_{active}$.  Our results are thus consistent with both a slow-acting quenching process that starts at infall and works continuously and a rapid quenching process that begins not at the time of infall, but after a delay period.

\section{Discussion }
\label{sec:discussion}

In this paper we have investigated the properties and incidence of star forming  galaxies in different environments, with the goal of helping constrain the physical mechanisms that  lead to galaxy quenching. Specifically, we have measured the differences between the SFR of each galaxy and the value derived from the field SFR-mass relation given the galaxy mass (\dsfr) and found it depends on both morphology, parametrized in terms of $B/T$, and environment (cluster core, infall and field). We have also quantified the fraction of galaxies with \dsfr$<$-0.45 (suppressed galaxies) in the different environments and at different  $B/T$ ratios. To try to disentangle the role of morphology and environment, we have first performed some comparisons at fixed $B/T$ and in the second part of the paper limited the analysis to galaxies with $B/T \le $0.3.  At a fixed $B/T$ the core and infall galaxies have systematically lower SFRs than the field (Fig.~\ref{fig:dsfr-bt}); these differences are statistically significant. Both considering the full range of $B/T$ and limiting the sample to disk-dominated galaxies, the fraction of galaxies with suppressed SFRs is higher in the core and infall regions relative to the field.  However, the properties of the suppressed galaxies (i.e. B/T, Sersic n, probability of being an Sc and color) are very similar in the different environments.  

Other works detected a population of cluster galaxies with reduced SFR compared to field galaxies of similar mass, both in the local universe \citep{Paccagnella2016} and at higher z 
\citep[][]{Patel2009, Vulcani2010, Guglielmo2019, Old2020}. Using spatially resolved data from  MaNGA, \citet{Belfiore2017} associated the suppressed population with a population of galaxies having central low ionisation emission-line regions, resulting from photoionisation by hot evolved stars, and star-forming outskirts. These galaxies are preferentially located in denser environments such as galaxy groups and are undergoing an inside-out quenching process. On the contrary, studies on galaxy samples based on a local parametrisation of environment do not find differences in the SFR–M of galaxies at different densities 
(\citealt{Peng2010, Wijesinghe2012}, but see also \citealt{Popesso2011} at high z). Also \citet{Cooper2022} do not detect a suppressed population at intermediate redshift, but their sensitivity limits are probably too high to detect the low-SFR population.

Our lack of difference between infall and core galaxies is not consistent with \citet{Paccagnella2016} though, who find more suppressed galaxies in the core than in the infall regions of the OMEGAWINGS clusters. Contamination from interloper galaxies will tend to reduce any observed differences between the core and infall samples and thus might partially explain the lack of difference we observe.  However, the \citet{Paccagnella2016} sample should also suffer from contamination from interlopers. The difference in results is not likely to be due to the different sample definitions, as the stellar mass completeness limit of the two works are comparable.  \citet{Paccagnella2016} do not limit their samples based on morphology, so part of the trends they find as a function of projected cluster-centric radius could be driven by the correlation between \dsfr \ and $B/T$ (Fig. \ref{fig:dsfr-bt}).  However, we do not detect a difference between the core and infall samples even when we use the full sample with all $B/T$ values. The SFRs used by \citet{Paccagnella2016} are derived from spectrophotometric models that are limited to the optical range, whereas the SFRs used in this study from from SED fitting that accounts for UV through IR emission.  However, it is not clear how this could be enough to reconcile our results. Regarding the properties of the cluster sample, \lcs \ contains only 9 clusters that span a range in mass, from massive groups to Coma, while the OMEGAWINGS sample contains 46 clusters whose masses are systematically higher ($\sigma_{cl}=$500-1100 km/s). Our differences could therefore be due to cluster-to-cluster variations and the more limited sample size used in this study.  One way forward is to complete a meta analysis of the combined \lcs \ and OMEGAWINGS samples, but this is not yet possible because the OMEGAWINGS data are not public.  Alternatively, one can construct a new cluster sample that is larger than both the LCS and OMEGAWINGS samples.  This would enable us to confirm trends with higher significance and to study variations among clusters. 
\citet{Paccagnella2016} is the only work that focuses on the most massive structures in the local universe, characterizes environment using halo mass, and analyzed the the SFR distribution of galaxies.  This is why we compare exclusively with this work.

Overall, it is  important to stress that different results in the literature obtained by adopting different parametrizations of environment are hard to compare, either because of the different selection criteria of the samples or custom definitions used to define, for example, the local galaxy over-density.

\subsection{Mechanisms Driving Quenching}

We can first consider the processes that could give rise to the  observed correlation between SFR with morphology,  specifically, that \dsfr \ decreases as \bt \ increases in all environments.  Multiple internal quenching mechanisms should scale with bulge size and could therefore give rise to the observed correlation. For example, AGN feedback should scale with \bt \ due to the bulge-black hole mass correlation \citep[e.g.][and references therein]{Beifiori2012, Kormendy2013}.  While we have removed AGN from the sample, we can not rule out their role in driving the \dsfr$-B/T$ correlation. 
 For example, radio-mode quenching could be happening with low-luminosity AGN that turn on and off.  If they are currently in the off phase, then we would not classify them as AGN, but
 they could be driving quenching.
 
 Other physical processes could provide a more direct link between the bulge and star formation quenching.  For example, bars and perhaps mergers that help build bulges will drive gas toward the center of galaxies, and this should lead to continued bulge growth and gas consumption.  
Alternatively, morphological quenching leads to reduced star-formation but not to a reduction in the gas content by stabilizing the disk gas against collapse \citep[][]{Martig2009}. 

We next consider processing that is driven by the large-scale environment.  Many physical processes can be invoked to explain the removal of gas in dense environments. In practice, it is extremely difficult to isolate a particular physical process, and it is more realistic to assume that many processes are working to deplete the gas in galaxies.  A further complication is that intrinsic phenomena can be enhanced in dense environments.  For example, the building of bulges in dense environments could drive an increase in morphological quenching or enhanced efficiency of radio AGN feedback from growing black holes.  
Nonetheless, our result that all $B/T$ galaxies have lower \dsfr \ in the cluster infall and core regions provides strong evidence in favor of additional environmentally-driven processing.
The presence of a population of suppressed galaxies can be be interpreted as evidence for a slow quenching process which prevents a sudden relocation of galaxies from the star forming to the red sequence
\citep[e.g.][]{Vulcani2010,Paccagnella2016}.
The lack of significant difference in the star-forming properties of the infall and core galaxies can be explained with a scenario where environmental processing is happening before the galaxies are accreted into the clusters \citep[e.g.][]{Zabludoff1998}.  Such pre-processing could result from starvation once a galaxy becomes a satellite \citep[][]{Larson1980}, or for galaxies that are accreted into a group prior to becoming cluster members, from galaxy-galaxy interactions that are prevalent in group environments.  Note that the latter could also contribute to bulge growth and perhaps the correlation between \dsfr \ and $B/T$.  Indeed, simulations show that half of $z=0$ cluster galaxies with low and intermediate stellar mass were first accreted into a group before entering the cluster \citep[e.g.][]{DeLucia2012}.  Alternatively, as we note in the Section \ref{sec:envs}, our infall sample could contain backsplash galaxies that have already traveled through the cluster core, and this would dilute any observed difference between our infall and core samples.  



\subsection{Timescale Associated with Environmentally-Driven Quenching}

In principle, the timescale associated with the decline in SFRs in dense environments can help identify the physical mechanisms that are driving quenching.
In practice, however, constraining the timescale is difficult.  For example, our observations are consistent with a quenching mechanism that acts continuously and over a long timescale, such as starvation \citep[e.g.][]{Larson1980}, and equally consistent with a scenario where the environmentally-induced decline of SFR starts after a delay period upon entering into the cluster, followed by a rapid period of quenching.  The latter delayed+rapid model is similar to what is proposed in \citet{Wetzel2012}, \citet{Haines2013}, and \citet{Rhee2020}, and this dependence of the inferred timescale on exactly how the quenching path is modeled is illustrated clearly in \citet{Cortese2021}.

The delay+rapid models from \citet{Wetzel2013} were originally devised to explain two observations: 1) that the quenching time inferred from the build-up of the passive population over cosmic time was long, and 2) that the distribution of SFRs, or more specifically the lack of suppressed SFRs seen in previous works \citep[e.g.][]{Peng2010}, indicated that the quenching time needed to be fast.  
Our modeling shares similarities to that from \citet{Wetzel2013} but there are important differences.  First, we control for morphology, which allows us to isolate changes in star formation from changes in morphology.  The importance of this component is demonstrated in Fig.~\ref{fig:dsfr-bt}.  Second, our sample extends to more massive clusters than \citet{Wetzel2013}.  Their sample has 160 ``groups" at $M_{vir}>10^{14}$M$_\odot$, and their most massive halo has  $M_{vir}=10^{15}$M$_\odot$.  Using the scaling from \citet{Finn2005} between cluster velocity dispersion and $M_{vir}$, we find that our sample has 4 clusters at $M_{vir}>10^{15}$M$_\odot$ and that all of our systems are at $M_{vir}>1.1\times 10^{14}$M$_\odot$. 

In comparing the actual quenching timescale constraints between our work and \citet{Wetzel2013}, we find some interesting differences.  {In \citet{Finn2018} we show that star-forming disk dominated galaxies in the cluster cores have smaller star-forming disks than galaxies in the infall regions. } This implies that the spatial distribution of star formation in galaxies is being affected by the environment even while they are star-forming and before any putative rapid quenching phase.  In addition, in this paper we have shown in our modeling that short delay times and long quenching times are allowed (Fig.~\ref{fig:sfr_pvalue}), {and this result is driven directly by our SFR distributions, specifically the distribution of galaxies with low SFRs.}  {This is in apparent conflict with \citet{Wetzel2013}, who conclude that long delay times are needed as their zero delay time model produces too many galaxies with intermediate sSFRs.}

{There are multiple possibilities to explain this disagreement.  First, we use SFRs based on UV through IR SED modeling \citep{Salim2018}, whereas the SFRs used by \citet{Wetzel2013} are derived from optical spectroscopy from \citet{Brinchmann04} with updated AGN and aperture bias corrections from \citet{Salim2007}.  \citet{Wetzel2013} use H$\alpha$-based SFRs down to $\rm \log_{10}(sSFR/yr^{-1}) = -11$, but below that they use a combination of emission lines. However, the UV provides more reliable SFRs for galaxies with low SFRs \citep{Salim2007}, and we may therefore be probing the low SFR end of the SFR distribution more accurately.  Second, we make different assumptions for how field galaxies appear prior to infall.  \citet{Wetzel2013} explicitly decompose their galaxies into satellites and centrals as a function of redshift using a subhalo abundance matching (SHAM) model.  We on the other hand use field galaxies in halos of $\log_{10} (M_{halo}/M_\odot) < 13$, of which $<10\%$ should be satellites at the time of first infall \citep{Wetzel2013}.  \citet{Wetzel2013} and our work both evolve our field galaxies back to the time of infall using the evolution of the SFR-stellar mass relation, though with different parameterizations of the relation \citep{Noeske2007,Whitaker2012}.  In calculating this evolution we take stellar mass loss into account in a more sophisticated way.
It is not clear which of these differences could result in the discrepancy between our different results, but we note that our zero delay model explicitly reproduces the distribution of SFRs over the same SFR range where the \citet{Wetzel2013} zero delay model does not.  Also, \citet{Wetzel2013} find a best-fit quenching timescale of 1.6--1.7~Gyr for their zero delay model, a time that we can rule out with our modeling (Fig.~\ref{fig:sfr_pvalue}).}  
The long quenching timescales with zero delay time that we find are consistent with the long quenching timescales required by directly modeling the build-up of the quenched fraction \citep[e.g.][]{McGee2009, DeLucia2007}.   

The connection between shrinking star-forming disks and suppressed star formation rates may be complicated.  For example, as discussed in \S\ref{sec:phase-space-disks} and Figure~\ref{fig:phasespace}, the phase space distribution of the normal and suppressed galaxies are similar, yet in \citet{Finn2018} we find that core galaxies have systematically smaller star-forming disks.  We will explore the connection between the size of the star-forming disk and the integrated SFR in a future paper.

\subsection{Estimating the Contamination from Passive Galaxies\label{sec:passive-contamination}}
As shown in Figure \ref{fig:ms_passive_cut}, there is a significant population of galaxies with very low inferred SFRs that fall below the cut we use to select our star-forming sample.  While our SFR limit is well above minimum reliable sSFR limit from \citet{Salim2018}, it may nonetheless be true that some of the galaxies in our suppressed region have their UV and re-radiated IR emission dominated by contributions from evolved stellar populations rather than from young massive stars.  That is, some of our suppressed galaxies may be scattered from the truly passive population and could contaminate our star-forming sample.  Furthermore, the fraction of passive galaxies in clusters is larger than in the field, and so contamination could contribute to the excess population of suppressed galaxies that we find in the clusters relative to the field.

To quantify the potential contamination from such passive galaxies, we use the double Gaussian fits discussed in {Appendix \ref{append:sfms_fit}} to model the passive population.  We assume the low-SFR peak is composed primarily of passive galaxies. 
We integrate the passive Gaussian from $-\infty$ to our passive cut (red line in Fig. \ref{fig:ms_passive_cut}) to get the number of passive field galaxies in a particular mass bin.
We assume that the distribution of passive galaxies in the field is the same as in the cluster, and we scale the field Gaussian to match the observed number of passive galaxies in the cluster ($N_{PCG}$), where the scale factor is:
\begin{equation}
\rm 
    scale = \frac{1}{N_{PCG}} \int_{-\infty}^{passive \ cut} \frac{A}{\sigma \sqrt{2\pi}} e^{\frac{-(SFR-\mu_{SFR})^2}{\sigma^2}} dSFR
\end{equation}
The width of the passive Gaussian is $\sigma$, the center of the passive peak is $\mu_{SFR}$, and the amplitude of the passive Gaussian is $A$.

To predict the number of passive galaxies that have SFRs in the region we define as suppressed, we integrate the scaled Gaussian from our passive cut to \dsfr$=-1.5 \sigma$, the upper bound that we use to define suppressed galaxies. {We again integrate from  \dsfr$=-1.5 \sigma$ to $10~\sigma$ to determine the number of passive galaxies that could contaminate our star-forming sample.}
The passive peak is not well defined at $\rm log_{10}(M_\star/M_\odot) < 10$ because the majority of field galaxies at these masses are forming stars.  We therefore limit the calculation of contamination to masses above this value.  After repeating the calculation in mass bins of 0.2~dex and summing the contribution from each mass bin, we find that there is a non-negligible number of passive galaxies that could be contaminating our star-forming population; {approximately 25\% of our suppressed galaxies could actually be passive galaxies.  If we correct for this contamination, as well as the contamination expected among our star-forming sample, the fraction of suppressed cluster galaxies decreases from 19\% to 15\%, whereas the fraction of suppressed field galaxies remains nearly constant at 7\%.  The significance of the difference between the cluster and field is reduced from $>4\sigma$ to $2.8\sigma$ after correcting for contamination.  Part of the reduced significance is due to the reduced sample size for the cluster that results from the subtraction of the potentially contaminating sources.  This test should be repeated using a larger cluster sample to better quantify the potential impact of contamination from passive galaxies.}

{We have presented one attempt to estimate the level of contamination from passive galaxies, but this method has limitations that likely overpredict the contamination.  Foremost, the contamination estimate depends heavily on the width of the Gaussian that we fit to the passive galaxies, yet it is not clear if a Gaussian is actually the appropriate model to use.  
On one hand, it is challenging, if not impossible, to discriminate the sSFRs of galaxies with very low sSFRs. While the distribution of these non-detections would likely be Gaussian,
the emission that drives the SFR measure for passive galaxies could also come from non-SFR related sources, e.g. thermally pulsating asymptotic giant branch stars \citep[TPAGB;][]{Kelson2010} or post asymptotic giant branch \citep[pAGB;][]{Salim2016} stars.  This would not necessarily result in a Gaussian distribution.  Furthermore, the presence of true suppressed or transition galaxies will force the center of the passive Gaussian to higher SFRs, and this will increase our estimate of the contamination.
We see the limitations of a Gaussian fit to the passive sequence in Figure~\ref{fig:double_gaussian}, where it is clear that the distribution of SFRs at log(SFR)$<-1$ is not Gaussian, especially for galaxies with log$(M_\star)>10$ where we compute the correction.  The result is an unphysical tail that extends to high SFRs.  Thus, our calculated contamination using this method is likely an overestimate.
Despite these limitations, 
upon applying these corrections we still detect an excess of suppressed galaxies in the clusters, although with somewhat reduced significance}.



\section{Summary and Conclusions}
\label{sec:conclusions}

In this work we have investigated the SFR-mass relation of galaxies residing in different environments in the local Universe. To characterize cluster galaxies, we made use of nine clusters from the {\it Local Cluster Survey} and considered separately core and infall galaxies. The field sample was instead drawn from the \cite{Tempel2014} catalog, and only galaxies belonging to halos with $\log (M_{halo}/M_\odot)<13$ were used. Galaxy properties (SFRs, stellar masses, structural parameters) were extracted from the same catalogs \citep{Simard2011, Salim2016, Salim2018} for both  samples, so that all quantities are homogeneous and can be directly compared. 
In our analysis, we  considered first all galaxies regardless of their morphology and then only galaxies with $B/T \le 0.3$, to investigate the role of morphology on the results.
Our main findings can be summarized as follows:
\begin{itemize}
    \item The SFR-mass relation depends on the environment: clusters (both in the core and in the infall regions) host a population of galaxies with lower SFRs -- but similar stellar masses --  than the field (Figure \ref{fig:lcsgsw-sfrmstar}). We define suppressed galaxies as those objects lying 1.5$\sigma$ below the fit of the field SFR-mass relation, and we observe an increase in the fraction of suppressed galaxies from the field, to the cluster infall and cores. (Figs. \ref{fig:dsfr-hist}, \ref{fig:fsuppressed}).
    
    \item Comparing the B/T and stellar mass distribution of suppressed and ``normal'' galaxies, we find that in all the environments galaxies with suppressed SFRs have higher B/T ratios than galaxies with normal SFRs, indicating that the  suppression of star-formation is linked to the growth of the  bulge for both field and cluster galaxies. 
    
    \item At any given B/T, cluster core and infall galaxies have a systematically lower SFRs than field galaxies, suggesting SFR is strongly linked to both the environment and morphology (Fig. \ref{fig:dsfr-bt}).  Our observational results thus
require a quenching mechanism that is linked to bulge growth that operates in all environments, and an additional mechanism that further reduces the gas content and SFRs of galaxies in dense environments.  
    
    \item The SFR-mass relation also depends on the environment when considering only disk-dominated galaxies ($B/T \le 0.3$) -- to limit the effect of morphology -- even though the population of suppressed galaxies is less evident. 
    \item A phase space analysis of the infall and core galaxies shows that suppressed and normal star-forming galaxies do not occupy distinct regions in phase space. 
    
\end{itemize}

We then model the SFRs of the disk-dominated core galaxies by creating a simulated core sample that is derived from the field.  We implement a model where accreted galaxies experience a delay phase during which no environmental processing occurs followed by a period of active quenching where SFRs decline exponentially.  We allow the delay phase to vary from zero to large values, and we compare the simulated and observed core galaxies at each combination of delay+active timescales.
Our results are consistent with both a slow-acting quenching process that starts at infall and works continuously and a rapid quenching process that begins after a delay period.  
This removes the necessity for a delay phase to explain the observed distribution of SFRs.


\section{Data Availability}

The data and python code used in this analysis are available at \url{https://github.com/rfinn/LCS-paper2}.

\section*{Acknowledgements}

The material is based upon work supported by NASA under
award No 80NSSC21K0640 and 80NSSC21K0641.
R.A.F. gratefully acknowledges support from NSF grants AST-0847430 and AST-1716657.
BV acknowledges financial contribution from the grant PRIN MIUR 2017 n.20173ML3WW\_001 (PI Cimatti) and  from the INAF main-stream funding programme (PI Vulcani).
G.H.R. acknowledges support from NSF-AST
1716690.
The authors thank the hospitality of the International Space
Science Institute (ISSI) in Bern (Switzerland) and of the
Lorentz Center in Leiden (Netherlands). Regular group meetings in these institutes allowed the authors to make substantial
progress on the project and finalize the present work.

This research made use of Astropy, a community
developed core Python package for Astronomy \citep{AstropyCollaboration2013,AstropyCollaboration2018}, matplotlib \citep{Hunter2007},
and TOPCAT \citep{Taylor2005}.



\bibliographystyle{mnras}
\bibliography{lcsms} 




\appendix

\section{Fitting the Star-Forming Main Sequence}
\label{append:sfms_fit}
\begin{figure*}
    \centering
    \includegraphics[trim = 0 100 0 0, width=0.9\textwidth]{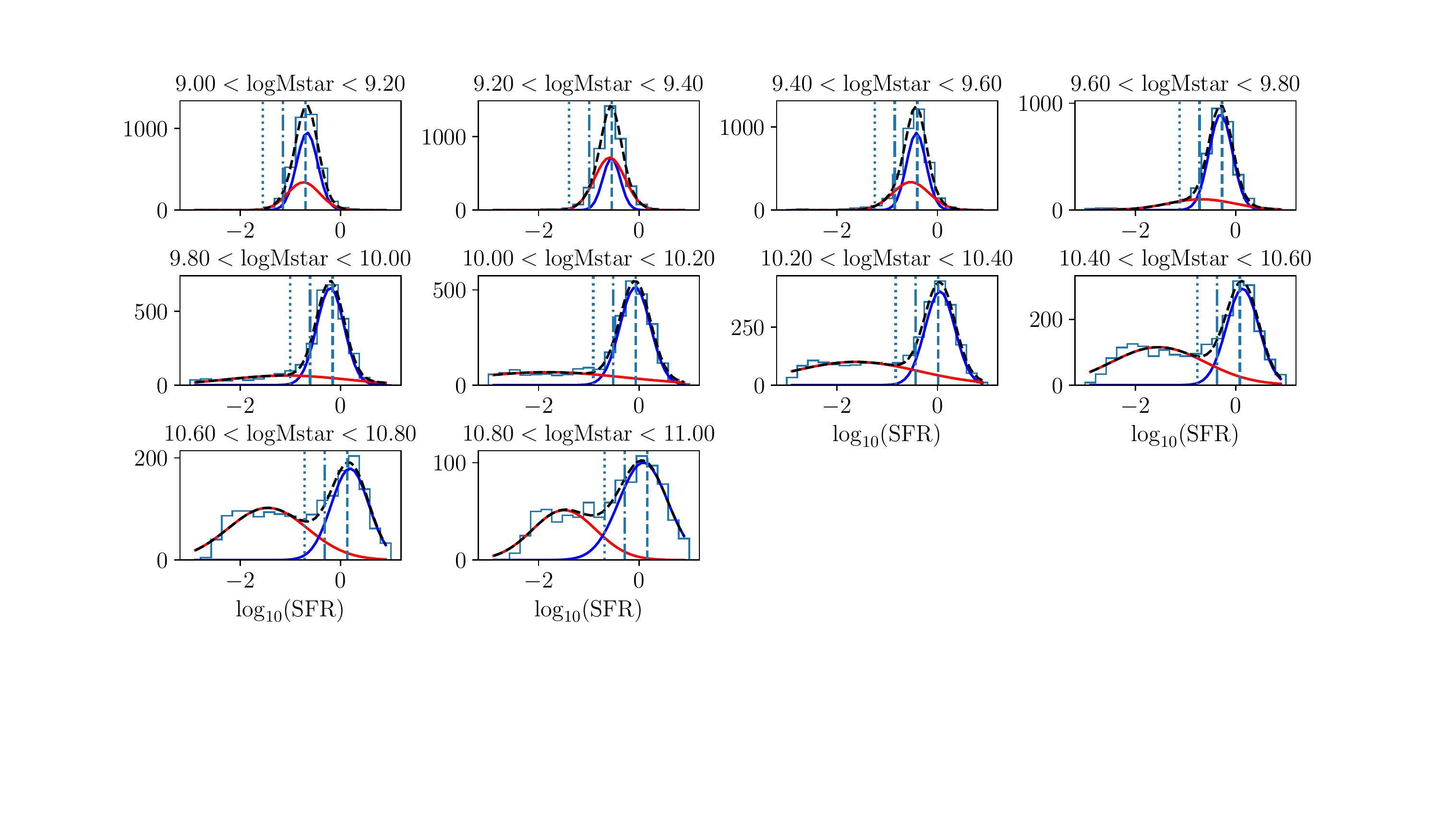}
    \caption{Histogram of SFRs for field galaxies, in mass bins of $\Delta m = 0.2$~dex.  We fit the histogram with a double gaussian, which we show with the black dashed line.  The blue peak shows the gaussian fit to the star-forming galaxies, and the red peak models the passive population.  The dashed vertical line shows the center of the main sequence, and the dot-dashed vertical line shows the main-sequence peak $-1.5\sigma = -0.45$~dex.  The cut we use to separate star-forming and passive galaxies is shown with the vertical cyan dotted line.}
    \label{fig:double_gaussian}
\end{figure*}

We fit the $SFR-M_\star$ relation using the $9 < \log_{10}(M_\star/M_\odot) < 11$ field galaxies.  We include galaxies with all $B/T$ values.   
To begin, the cyan histograms in Figure \ref{fig:double_gaussian} show the distribution of SFRs in mass bins of 0.2~dex.  The distribution is single-peaked at low mass and double-peaked at higher masses.  The primary peak is due to star-forming galaxies, and the flatter secondary peak that grows more prominent at $\log_{10}(M_\star/M_\odot) > 10$ includes both transition and passive galaxies.  To isolate the properties of the star-forming galaxies, we fit a double Gaussian to SFR distribution.  We show the results in Figure \ref{fig:double_gaussian}.  The blue curve is the fit to the star-forming peak, the red curve is the fit to the passive peak, and the dashed black curve is the sum. We note that fitting a Gaussian to the passive sequence is not a proper representation of the true distribution of SFRs for passive galaxies.  Rather, we adopt this technique as an attempt to deal with the important issue of how the scatter in passive galaxies might contribute to the low-SFR tail of the SFR distribution.  This contamination is relevant both to the fit to the main sequence and also to the contamination of passive galaxies into the low-SFR end of our star-forming galaxy population.  We have fit the main sequence both using the star-forming peak from the double-Gaussian fit and also using all galaxies above our sSFR limit.  Our results are unchanged if we adopt the main sequence fit to the entire sample.

To characterize the SFR as a function of stellar mass, 
we fit the center of the star-forming peak versus stellar mass.  Specifically, we fit in log-log space using a second-order polynomial:  
\begin{equation}
\log_{10}\left(SFR_{M_{\sun}~yr^{-1}}\right)= a \log_{10}(M_{\star}//M_{\sun})^2 + b \log_{10}(M_{\star}//{M_{\sun}}) + c.
\label{eqn:sfms}
\end{equation}
The best-fit coefficients are: 
$a = -0.197 \pm 0.003$, $b=4.42 \pm 1.23$, and $c=-24.6\pm30.6$.
The scatter in the relation is 
 $\sigma = 0.3$~dex.  

In part of the analysis, we analyze the star-formation properties of disk-dominated galaxies, and we refit the main sequence for the $B/T \le 0.3$ sample using a similar procedure.  However, the SFR distribution does not exhibit a strong bimodal behavior at any masses when considering $B/T \le 0.3$ field galaxies only.  Thus, we fit a single Gaussian to the SFR distribution.  As before, we fit a second order polynomial to the peak of the SFR distribution as a function of stellar mass (see Eqn. \ref{eqn:sfms}; best fit coefficients: $a=-0.0949\pm0.0004$, $b=2.459\pm 0.168$, and $c = -15.23 \pm 4.19$ ). The standard deviation of the main sequence is still 0.3~dex for the $B/T \le 0.3$ field sample.

\section{Separating Passive and Star-Forming Galaxies }
\label{append:passive}
While \citet{Salim2018} provide a minimum reliable sSFR of $\rm \log_{10}(sSFR/yr^{-1}) = -11.5$, some galaxies with this formal specific SFR may in fact be passive galaxies.  We adopt a more conservative SFR cut to reduce the passive contamination.  We first determine the 95th percentile of the field mass distribution, $\log_{10}(M_\star/M_\odot) = 10.8$.  We then determine the shift required to bring the main sequence down to the  $\rm \log_{10}(sSFR/yr^{-1}) = -11.5$ limit of the GSWLC-2 survey at this mass value.  We show the result with the red line in Figure \ref{fig:ms_passive_cut}.  This line is parallel to the main-sequence fit but shifted 0.845 dex below it.

\section{SFR-Stellar Mass Relation and Passive Cut for Disk Dominated Galaxies}
\label{append:disk}
\begin{figure*}
    \centering
    \includegraphics[width=.45\textwidth]{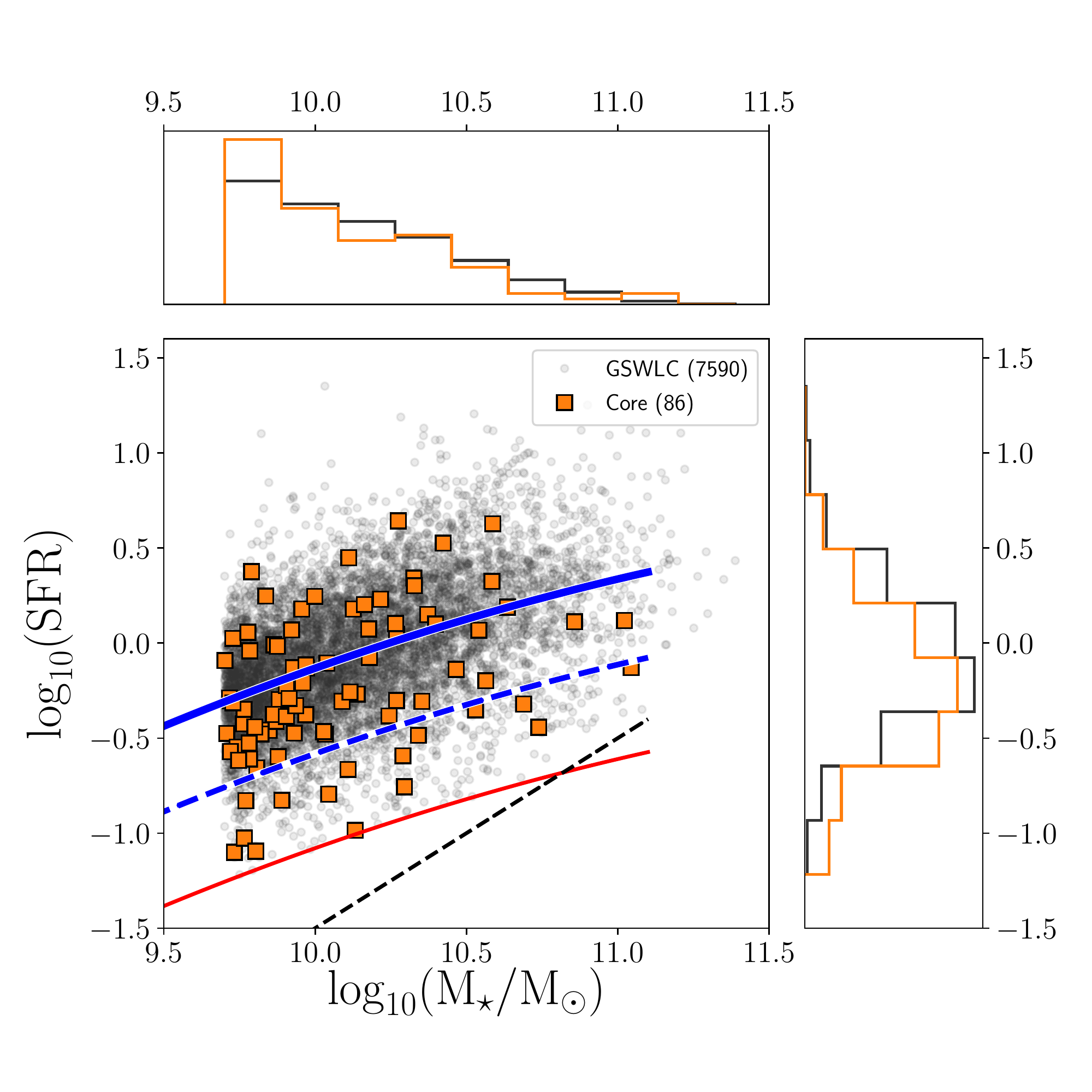}
    \includegraphics[width=.45\textwidth]{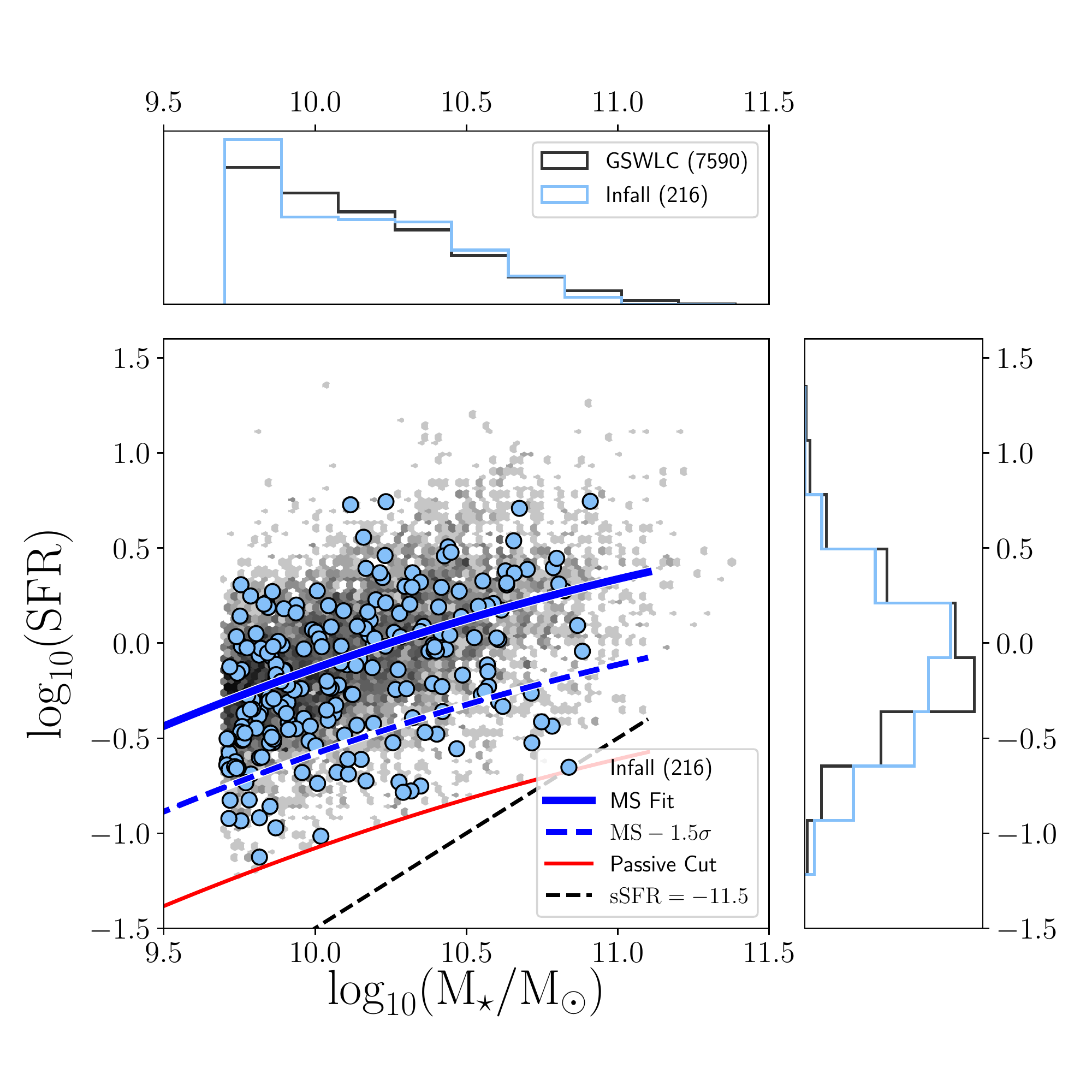}
    \caption{(Left) $\log_{10} (SFR)$  versus $\log_{10}(M_\star/M_\odot)$ for LCS core (orange) and field galaxies (gray).  
    Only $B/T \le 0.3$ galaxies are included.  
    The two samples have a similar distribution of stellar masses, but the LCS core galaxies have a larger fraction of suppressed SFR galaxies.  The solid blue line shows our fit to the $SFR-M_\star$ relation, and the dashed blue line shows the fit minus $1.5\sigma$.  The red line is the division that we adopt to separate star-forming and passive galaxies.  The dashed black line is the specific SFR limit of $sSFR = -11.5$.   (Right) Same as left panel but for LCS infall and field galaxies.  A KS test is not able to distinguish the stellar mass and SFR distributions for LCS infall (light blue) and field galaxies (gray).  
    }
    \label{fig:lcsgsw-sfrmstar_BTcut}
\end{figure*}

Similarly to what has been done for the full sample in Sec. \ref{sec:galaxy_properties}, for the analysis presented in Sec. \ref{sec:disk_restrict} onward, we fit the main sequence fitting for the $B/T \le 0.3$ sample.  We set the lower SFR limit using a curve that is parallel to the main sequence and intersects the sSFR limit of the GSWLC-2 ($sSFR = -11.5$) at the 95th percentile of the mass distribution ($\rm \log_{10}(M_\star/M_\odot) = 10.8$).  
The division between the star-forming and passive $B/T \le 0.3$ galaxies 
is parallel to the main-sequence fit but shifted 0.947 dex below it.  
We show the SFR-stellar mass relation and the main sequence fits for $B/T \le 0.3$ galaxies in Figure \ref{fig:lcsgsw-sfrmstar_BTcut}. This fit is similar to the main sequence fit for the full sample, except that the full sample exhibits more curvature at $\rm \log_{10}(M_\star/M_\odot) = 10.5$. 

Similarly to what found for the full sample, the SFR distribution of infall/core sample galaxies is skewed towards lower values than the distribution of the field sample, while the mass distributions are more similar.

\section{Mass Completeness Limit}
\label{append:mass_completeness}
We estimate the stellar mass completeness limit of our sample to be $\log_{10} (M_\star/M_\odot) = 9.7$.
This limit reflects the minimum mass above which we can detect galaxies regardless of their $r$-band stellar mass-to-light ratios (\mlstar).  As is always the case with determining stellar mass limits from magnitude limited data, we cannot determine the stellar mass limit using the lowest mass of our faintest galaxies, as those low mass objects will be biased towards the lowest \mlstar.  In other words, we would be missing galaxies at our magnitude limit at the same mass, but with higher \mlstar.  Instead, we determine the \mlstar\ limit using a technique adapted from that in \citet{Marchesini2009} and \citet{Rudnick2017}, in which we use galaxies in our own sample that are at the redshift of our highest redshift cluster ($z\approx 0.04$), are above the GSWLC sSFR limit of -11.5, and are approximately one magnitude brighter than the GSWLC2 magnitude limit of $r = 17.7$.  This bright subsample is far enough above the magnitude completeness limit so as to be equally complete for all $r$-band stellar mass-to-light ratios.  Under the reasonable assumption that the \mlstar\ distribution does not vary strongly over 1 $r$-band magnitude, this bright subsample should reflect the intrinsic \mlstar\ distribution of galaxies near our apparent magnitude cut.   We then use this brighter subsample to estimate how massive a galaxy at our apparent magnitude limit needs to be in order to be detected regardless of its \mlstar.  We proceed by calculating the amount of fading required to bring each galaxy in the bright subsample to the magnitude limit of $r=17.7$.  We then decrease the stellar masses by the same factor.  For example, a galaxy which was one magnitude brighter than the $r=17.7$ limit would have its stellar mass reduced by 0.4~dex.
The resulting distribution of shifted stellar masses has by construction identical apparent magnitudes and luminosities.  This faded sample represents the intrinsic distribution in \mlstar\ (and hence in $M_\star$) for galaxies at our magnitude limit, but unaffected by that same limit.  We select the highest 5\% in $M_\star$ of this shifted sample as our stellar mass completeness limit, above which we are complete for all galaxies regardless of their \mlstar.  Galaxies at lower stellar masses, and therefore lower \mlstar, would be detected at our $r$-band magnitude limit, but only if they had lower \mlstar\ values.  Such galaxies would therefore be biased against objects with, e.g. fainter stellar populations or suppressed star formation.


\bsp	
\label{lastpage}
\end{document}

%% file: table1-massmatch.tex
\begin{table*}
\centering 
\begin{tabular}{|c|c|c|c|} 
\hline
Samples  & Variable  &\multicolumn{1}{c|}{All $B/T$} &  \multicolumn{1}{c|}{$B/T < 0.3$}\\ 
& &  {A.D. p value}& {A.D. p value} \\ 
\hline \hline 
Core-Field     & $\log SFR$         &  1.21e-03 &  1.00e-03  \\ 
 & $\Delta \log $SFR  &  1.00e-03 &  1.00e-03  \\ 
 & $B/T$    & 3.39e-02 &  1.00e-03   \\ 
\hline 
Infall-Field  & $\log$ SFR & 1.00e-03 &  1.00e-03\\ 
& $\Delta \log$SFR  &  1.00e-03   &  1.00e-03 \\ 
&$B/T$   &   1.00e-03 &  4.03e-03  \\ 
\hline 
Core-Infall   & $\log$ SFR      &  2.50e-01    &  2.50e-01   \\ 
& $\Delta \log$SFR&  2.50e-01 &  2.50e-01  \\ 
& $B/T$ & 2.50e-01  & 1.19e-01   \\ 
\hline 
\end{tabular} 
\caption{Summary statistics for SFR, $\Delta$SFR,  and $B/T$.  Populations are significantly different when the Anderson-Darling p-value$<$3.0e-03. NOTE: {\tt scipy.stats.anderson\_ksamp} floors the p value at 0.1\% and caps the p value at 25\%, so it will not return p values below 1E-3 or above 0.25.} 
\label{tab:stats} 
\end{table*} 